\documentclass[sigconf]{acmart}

\usepackage{multirow}
\usepackage{enumitem}
\usepackage{pifont}
\usepackage{subfigure}
\usepackage{cleveref}
\crefformat{section}{\S#2#1#3}
\crefformat{subsection}{\S#2#1#3}
\crefformat{subsubsection}{\S#2#1#3}
\usepackage{balance}

\copyrightyear{2020}
\acmYear{2020}
\setcopyright{acmcopyright}\acmConference[PACT '20]{Proceedings of the 2020 International Conference on Parallel Architectures and Compilation Techniques}{October 3--7, 2020}{Virtual Event, GA, USA}
\acmBooktitle{Proceedings of the 2020 International Conference on Parallel Architectures and Compilation Techniques (PACT '20), October 3--7, 2020, Virtual Event, GA, USA}
\acmPrice{15.00}
\acmDOI{10.1145/3410463.3414626}
\acmISBN{978-1-4503-8075-1/20/10}

\begin{document}
\fancyhead{}

\title{Helix: Algorithm/Architecture Co-design for Accelerating Nanopore Genome Base-calling}

\author{Qian Lou}
\email{louqian@iu.edu}
\affiliation{%
\institution{Indiana University Bloomington}
}
\author{Sarath Janga}
\email{scjanga@iupui.edu}
\affiliation{%
\institution{Indiana University - Purdue University Indianapolis}
}
\author{Lei Jiang}
\email{jiang60@iu.edu}
\affiliation{%
\institution{Indiana University Bloomington}
}

\begin{abstract}
Nanopore genome sequencing is the key to enabling personalized medicine, global food security, and virus surveillance. The state-of-the-art base-callers adopt deep neural networks (DNNs) to translate electrical signals generated by nanopore sequencers to digital DNA symbols. A DNN-based base-caller consumes $44.5\%$ of total execution time of a nanopore sequencing pipeline. However, it is difficult to quantize a base-caller and build a power-efficient processing-in-memory (PIM) to run the quantized base-caller. Although conventional network quantization techniques reduce the computing overhead of a base-caller by replacing floating-point multiply-accumulations by cheaper fixed-point operations, it significantly increases the number of systematic errors that cannot be corrected by read votes. The power density of prior nonvolatile memory (NVM)-based PIMs has already exceeded memory thermal tolerance even with active heat sinks, because their power efficiency is severely limited by analog-to-digital converters (ADC). Finally, Connectionist Temporal Classification (CTC) decoding and read voting cost 53.7\% of total execution time in a quantized base-caller, and thus became its new bottleneck.

In this paper, we propose a novel algorithm/architecture co-designed PIM, Helix, to power-efficiently and accurately accelerate nanopore base-calling. From algorithm perspective, we present systematic error aware training to minimize the number of systematic errors in a quantized base-caller. From architecture perspective, we propose a low-power SOT-MRAM-based ADC array to process analog-to-digital conversion operations and improve power efficiency of prior DNN PIMs. Moreover, we revised a traditional NVM-based dot-product engine to accelerate CTC decoding operations, and create a SOT-MRAM binary comparator array to process read voting. Compared to state-of-the-art PIMs, Helix improves base-calling throughput by $6\times$, throughput per Watt by $11.9\times$ and per $mm^2$ by $7.5\times$ without degrading base-calling accuracy.
\end{abstract}



\maketitle

\section{Introduction}
\label{s:intro}

Genome sequencing~\cite{Turakhia:ASPLOS2018,Wu:HPCA2019,Turakhia:HPCA2019,Madhavan:ISCA2014,Fuijiki:ISCA2018} is a cornerstone for enabling personalized medicine, global food security, and virus surveillance. The emerging nanopore genome sequencing technology~\cite{Jain:Nature2018} is revolutionizing the genome research, industry and market due to its ability to generate ultra-long DNA fragments, aka \textbf{long reads}, as well as provide portability. Producing long reads~\cite{Nakano:Human2017} is the key to improving the quality of \emph{de novo} assembly, spanning repetitive genomic regions, and identifying large structural variations. Moreover, portable real-time USB Flash drive size nanopore sequencers, MinION~\cite{Jain:Nature2018} and SmidgION~\cite{ONT:SmidgION2020}, have demonstrated their power in tracking genomes of Ebola~\cite{Hoenen:EID2016}, Zika~\cite{Faria:GM2016} and COVID-19~\cite{Lu:Lancet2020} viruses during disease outbreaks.

Compared to conventional short-read Illumina sequencing, nano-pore sequencing suffers high error rate~\cite{Jain:Nature2018}, e.g., 12\%. A nanopore sequencer measures changes in electrical current as organic DNA fragments pass through its pore. Due to the tiny amplitude of currents triggered by DNA motions, a nanopore sequencer inevitably introduces noises into raw electrical signals, thus producing sequencing errors. A base-caller translates raw electrical signals to digital DNA symbols, i.e., $[A, C, G, T]$. In order to reduce sequencing errors, a sequencing machine generates multiple unique reads~\cite{Jain:Nature2018} that include a given DNA symbol. These reads are base-called individually, and then assembled to decide the correct value of each DNA symbol. The number of unique reads containing a given DNA symbol is called coverage. Typically, the coverage is between $30\sim 50$~\cite{Teng:GigaScience2018,Wick:GB2019,Scrappie:ONT2019}. To further enhance base-calling accuracy, recent works~\cite{Bovza:PloS2017,Teng:GigaScience2018,Wick:GB2019,Scrappie:ONT2019,Flappie:ONT2019} use deep neural networks (DNNs) for base-calling. A DNN-based base-caller, e.g., Guppy~\cite{Wick:GB2019}, Scrappie~\cite{Scrappie:ONT2019}, and Chiron~\cite{Teng:GigaScience2018}, consists of convolutional, recurrent, fully-connected layers, as well as a Connectionist Temporal Classification (CTC) decoder. Although achieving high base-calling accuracy, prior DNN-based base-callers are slow. For instance, Guppy with its high base-calling accuracy obtains only 1 million base pairs per second (bp/s) on a server-level GPU. \textit{At such a speed, it takes 25 hours for Guppy to base-call a 3G-bp human genome with a $30\times$ coverage}. During virus outbreaks, it is challenging for even a data center equipped with powerful GPUs to processing base-calling for a large group of presumptive positive patients. As a result, base-calling becomes the most time-consuming step in a nanopore sequencing pipeline~\cite{Cali:Brefings2018}.

Recently, both industry~\cite{Lin:ICML2016} and academia~\cite{Li:CVPR2019,Xu:ICLR2018} proposed network quantization algorithms to power-efficiently accelerate DNN inferences without sacrificing inference accuracy by approximating inputs, weights and activations of a DNN to fixed-point representations with smaller bit-widths. In this way, computationally expensive floating-point multiply-accumulates (MACs) in a DNN can be replaced by fixed-point operations. Besides conventional CPUs and GPUs, FPGAs and ASICs are adopted to accelerate quantized DNN inferences in data centers. Moreover, to further overcome the \textit{von Neumann} bottleneck in data centers, recent search efforts use various nonvolatile memory (NVM) technologies including ReRAM~\cite{Shafiee:ISCA2016,Yang:ISCA2019}, PCM~\cite{Ambrogio:IEDM2019} and STT-MRAM~\cite{Yan:ICS2018} to build processing-in-memory (PIM) accelerators to process quantized DNN inferences in memory arrays.

However, it is difficult to apply prior network quantization techniques on base-callers and accelerate quantized base-callers by state-of-the-art NVM PIM architectures. Na\"{\i}vely quantizing a base-caller via prior network quantization algorithms substantially increases the number of \textit{systematic} errors that cannot be corrected by voting operations among multiple reads containing the same DNA symbols. Furthermore, state-of-the-art PIM accelerators take advantage of analog computing to maximize inference throughput of quantized DNNs, but the functioning of their analog computing style heavily depends on a large number of CMOS analog-to-digital converters (ADCs) that significantly increase their power consumption and area overhead. For instance, CMOS ADCs cost 58\% of power consumption and 30\% of chip area in a typical PIM design~\cite{Shafiee:ISCA2016}. Finally, state-of-the-art NVM PIM designs cannot process some essential operations of a base-caller such as CTC decoding and read voting that usually consume $>$50\% of total execution time in a quantized base-caller.

\begin{figure}[t!]
\centering
\includegraphics[width=3.4in]{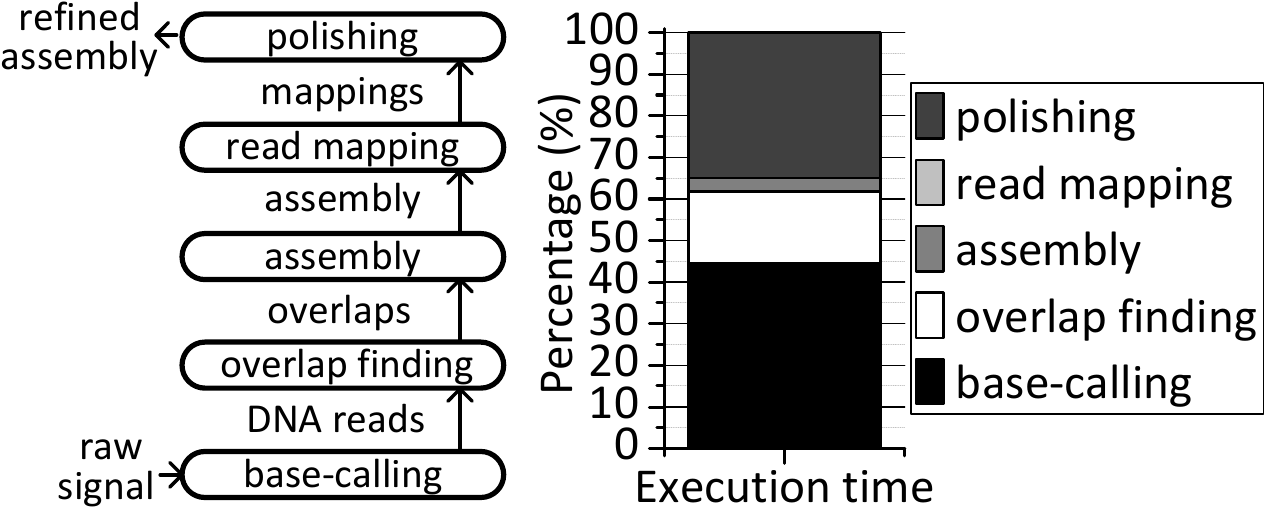}
\caption{The pipeline of nanopore sequencing.}
\label{f:dna_nano_pipeline}
\vspace{-0.2in}
\end{figure}

In this paper, we propose a novel algorithm and architecture co-designed PIM accelerator, \textit{Helix}, to efficiently and accurately process quantized nanopore base-calling. Our contributions are summarized as:
\begin{itemize}[noitemsep,topsep=0pt,leftmargin=*]
\item \textbf{Systematic error aware training}. 
We present systematic error aware training (SEAT) to reduce the number of systematic errors that cannot corrected by read votes in a quantized base-caller. We introduce a new loss function to indirectly minimize the edit distance between a consensus read and its ground truth DNA sequence. SEAT enables 5-bit quantized base-callers to achieving their full-precision base-calling accuracy.

\item \textbf{An ADC-free PIM accelerator}.
We propose a Spin Orbit Torque MRAM (SOT-MRAM)-based array architecture to accelerate analog-to-digital conversion operations without CMOS ADCs. We also show our SOT-MRAM ADC arrays are resilient to process variation. We modify a conventional NVM-based dot-product engine to accelerate CTC decoding operations, and then present a SOT-MRAM-based binary comparator array to process read voting operations in a quantized base-caller.

\item \textbf{Base-calling accuracy and throughput}.
We implemented all proposed techniques of Helix and compared Helix against state-of-the-art PIM designs that accelerate quantized DNN inferences. Experimental results show that, compared to state-of-the-art PIM accelerators, Helix improves base-calling throughput by $28\times$, throughput per Watt by $80\times$, and throughput per $mm^2$ by $27\times$ without degrading accuracy.
\end{itemize}

\section{Background}
\label{s:background}

\subsection{Nanopore Sequencing Pipeline}

As Figure~\ref{f:dna_nano_pipeline} shows, a nanopore sequencing pipeline~\cite{Cali:Brefings2018} consisting of \textit{base-calling}, \textit{overlap finding}, \textit{assembly}, \textit{read mapping}, and \textit{polishing} is employed to generate a digital assembly. The input of a pipeline is raw electrical signals produced by nanopore sequencers, e.g., MinION~\cite{Jain:Nature2018} and SmidgION~\cite{ONT:SmidgION2020}. Base-calling translates raw signal data to digital DNA symbols, i.e., $[A, C, G, T]$. Overlap finding computes all suffix-prefix matches between each pair of reads, and then generates an overlap graph, where each node denotes a read and each edge indicates the suffix-prefix match between two nodes. The assembly step traverses an overlap graph to construct a draft assembly. Base-called reads are mapped to the generated draft assembly by read mapping. Lastly, the final assembly is polished. 

\begin{figure}[t!]
  \centering
  \begin{minipage}{.61\linewidth}
    \centering
    \includegraphics[width=1.9in]{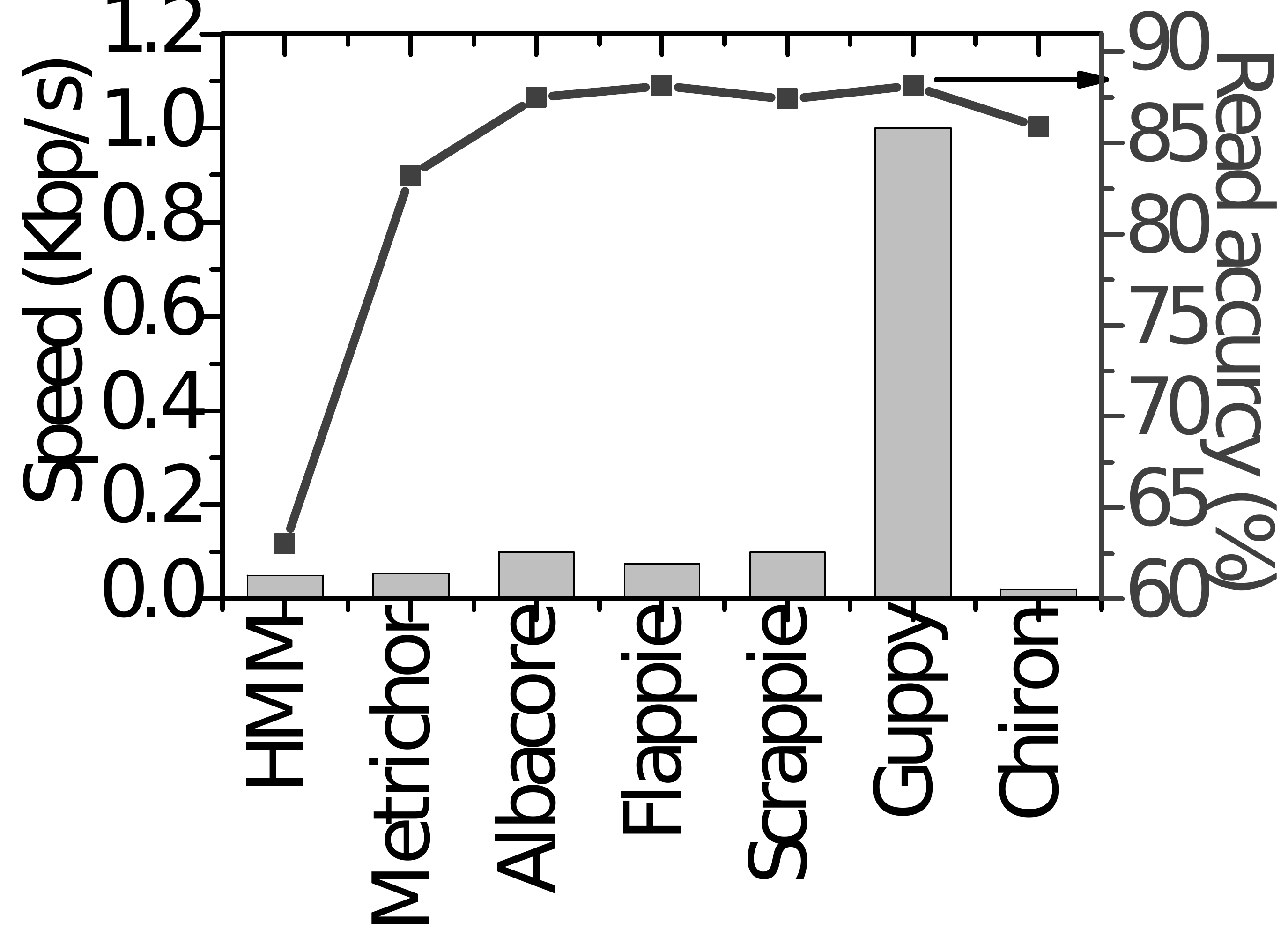}
		\vspace{-0.1in}
    \caption{Base-caller comparison.}
		\label{f:dna_basecaller_all}
  \end{minipage}
  \begin{minipage}{.39\linewidth}
    \centering
		\includegraphics[width=1.2in]{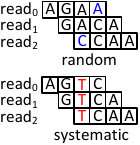}
    \caption{Errors.}
		\label{f:dna_basecalling_error}
  \end{minipage}
\vspace{-0.2in}
\end{figure}

\subsection{Nanopore Base-calling}

\textbf{DNN-based base-caller}. DNNs are adopted to filter noises and accurately translate raw electric signals to digital DNA symbols. A DNN-based base-caller typically consists of multiple convolutional (Conv), gated recurrent unit (GRU), and fully-connected (FC) layers. The convolutional layers recognize local patterns in input signals, whereas the GRU layers integrate these patterns into base-calling probabilities. A CTC decoder is used to compute digital DNA symbols according to the base probabilities. Compared to the Hidden Markov Model (HMM)~\cite{Metrichor:ONT2017}, a series of DNN-based base-callers including Metrichor~\cite{Metrichor:Online2018}, Albacore~\cite{Oxford:Albacore2018}, Flappie~\cite{Flappie:ONT2019}, Scrappie~\cite{Scrappie:ONT2019}, Guppy~\cite{Wick:GB2019}, and Chiron~\cite{Teng:GigaScience2018}, significantly improve base-calling accuracy, as shown in Figure~\ref{f:dna_basecaller_all}. Among all base-callers, the Oxford Nanopore Technologies official GPU-based base-caller, Guppy, achieves the best accuracy and the highest speed. We selected Guppy as our base-caller baseline, and also considered other DNN-based base-callers in~\cref{s:eval_analy}. Due to complex DNN structures, base-callers are generally slow~\cite{Wick:GB2019}. As a result, base-calling consumes 44.5\%~\cite{Cali:Brefings2018} of total execution time of a nanopore sequencing pipeline. The details of base-callers are introduced in~\cref{s:dna_other_callers}.

\textbf{Base-calling error}. We define the number of base-calling errors as the edit distance between a read predicted by a base-caller and its ground truth. The edit distance quantifies how dissimilar two reads are to one another by counting the minimum number of insertions, deletions, and substitutions required to transform one into the other. To enhance base-calling accuracy, a base-caller translates each signal data multiple times and generates multiple reads containing the same signal data. At the end of base-calling, each DNA symbol value is decided by votes among all reads containing its corresponding signal data. As Figure~\ref{f:dna_basecalling_error} shows, for a DNA symbol, if base-calling errors randomly occur among reads, the voting result can still be correct, since most reads have the correct value. This is a \textit{random} error. However, for a DNA symbol, if base-calling errors happen in a systematic way, i.e., all copies of a signal are translated to the same wrong value, it is impossible to produce the correct value by read voting. It is a \textit{systematic} error.

\begin{figure}[t!]
\vspace{0.1in}
\centering
\includegraphics[width=3in]{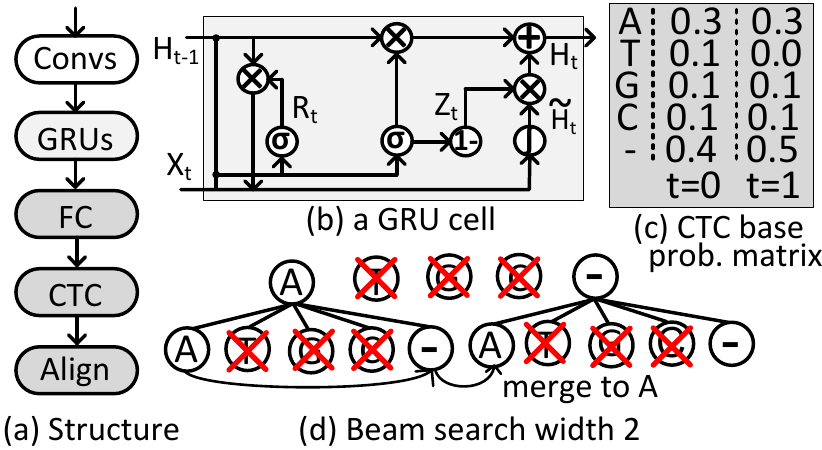}
\vspace{-0.1in}
\caption{The DNN architecture of Guppy.}
\label{f:dna_guppy_arch}
\vspace{-0.1in}
\end{figure}

\begin{figure}[t!]
  \centering
  \begin{minipage}{.61\linewidth}
    \centering
    \includegraphics[width=1.8in]{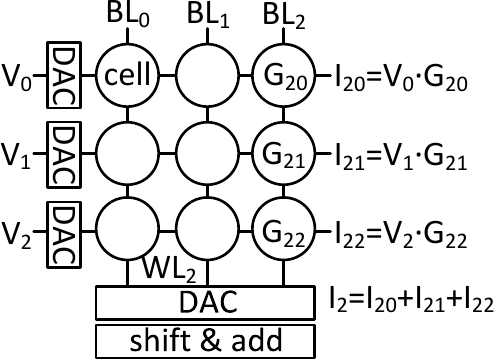}
	  \vspace{-0.1in}
    \caption{A dot-product engine.}
		\label{f:dna_dot_product}
  \end{minipage}\hspace{-0.15in}
  \begin{minipage}{.39\linewidth}
    \centering
		\includegraphics[width=1in]{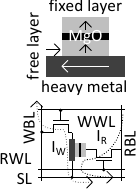}
		\vspace{-0.1in}
    \caption{SOT-MRAM.}
		\label{f:stt_sot_cell}
  \end{minipage}
\vspace{-0.2in}
\end{figure}

\textbf{Convolutional layer}. As Figure~\ref{f:dna_guppy_arch}a shows, a base-caller includes multiple convolutional layers to process raw electric signals. The first convolutional layer receives an $L\times N$ floating-point signal vector, where $L$ is the input length; and $N$ indicates the input channel number, e.g., $L=5$ and $N=1$. Then, it uses a $K\times N\times M$ weight filter to convolve with the input vector to generate an output vector for the next activation layer~\cite{Teng:GigaScience2018}, where $K$ is the weight kernel size; and $M$ means the output channel number, e.g., $K=2$, and $M=256$. The $L\times N$ floating-point signal vector is generated by a fixed-size window sliding on the entire signal data array. After a base-calling operation, the sliding window moves forward by $T$ elements~\cite{Teng:GigaScience2018}, where $T$ is the sliding offset, e.g., $T=1$. The base-caller then works on a new signal vector. At the end of base-calling, $\lfloor L/T \rfloor$ reads containing the same signal element vote for its value.

\textbf{GRU Layer}. A base-caller uses a set of GRU layers to integrate patterns produced by convolutional layers into base-calling probabilities. As Figure~\ref{f:dna_guppy_arch}b describes, a GRU layer receives an input $X_t$ and its output of the last time step $H_{t-1}$. And then, it uses two memory cells, $R_t$ and $Z_t$, to reset and update the gate state at the time step $t$. The output $H_t$ of a GRU layer can be computed as
\begin{equation}
\begin{aligned} 
Z_{t} &= \sigma (W_{z}X_t + U_{z}H_{t-1}) + b_z\\
R_{t} &= \sigma (W_{r}X_t + U_{r}H_{t-1})+b_r\\
\tilde{H_{t}} &= \smallint (W_{h}X_t + U_{h}(R_t \otimes H_{t-1})) + b_h\\
{H_{t}} &= Z_{t}\otimes H_{t-1} + (1-Z_{t}) \otimes \tilde{H_{t}}
\end{aligned}
\end{equation}
where $W_z$, $U_z$, $W_r$, $U_r$, $W_h$ and $U_h$ are weights for $Z_t$, $R_t$ and hidden state $\tilde{H_{t}}$ respectively; $b_z$, $b_r$ and $b_h$ are their biases; $\sigma$ is the \emph{sigmoid} activation; $\smallint$ indicates the \emph{tanh} activation; and $\otimes$ means element-wise multiplications.

\textbf{CTC decoder}. Since it is difficult for a nanopore sequencer to precisely control DNA motions at uniform speed, multiple elements in the input signal vector may be generated by a single DNA nucleotide~\cite{Teng:GigaScience2018}. A base-caller adopts a CTC decoder~\cite{Graves:ICML2006,Hannun:DISTILL2017} to map an input signal vector $R=[I_0, I_1, \ldots, I_{L-1}]$ to a corresponding digital read $D=[H_0, H_1, \ldots, H_{Z-1}]$, where $L\neq Z$; and there is no alignment between $R$ and $D$. More specifically, convolutional, GRU and FC layers provide all symbol probabilities $p_t(a_t|R)$ for each time step, where $a_t\in [A, C, G, T, -]$ ($-$ indicates blank). The probabilities $p_t(a_t|R)$ of a symbol of all time steps form a base probability matrix, as shown in Figure~\ref{f:dna_guppy_arch}c. By looking up the base probability matrix, a CTC decoder can decide the probability of a read. The probability of $D$ is calculated by 
\begin{equation}
p(D|R)=\sum_{A\in \mathbb{A}_{D,R}} \prod_{t=0}^{L-1} p_t(a_t|R)
\label{e:dna_prob_cal}
\end{equation}
where $\mathbb{A}_{D,R}$ indicates all valid alignments between $D$ and $R$. The CTC decoder infers the most likely read by a \textit{beam} search on the matrix. As Figure~\ref{f:dna_guppy_arch}d highlights, during a beam search with width 2, the CTC decoder keeps only the symbols with the top-2 largest probabilities at each time step. At $t=0$, it keeps $A$ and $-$. At $t=1$, the decoder calculates the probabilities for various 2-symbol reads including $p(AA)=0.3*0.3=0.09$, $p(A-)=0.15$, $p(-A)=0.12$, and $p(--)=0.2$. Since $AA$, $A-$, $-A$ indicate $A$, they can be merged to $A$. So $p(A)=0.09+0.15+0.12=0.36$. The beam search finds $A$ as the most likely read.

\subsection{Network Quantization}
To reduce the computing overhead of DNNs, recent work proposes network quantization~\cite{Lin:ICML2016,Li:CVPR2019,Xu:ICLR2018} that approximates 32-bit floating-point inputs, weights and activations to their fixed-point representations with smaller bit-widths. In this way, the quantized networks perform quantized inferences by low-cost fixed-point MACs.

\subsection{NVM-based Dot-Product Engine}

Various NVM-based dot-product engines (e.g., STT-MRAM~\cite{Yan:ICS2018}, PCM~\cite{Ambrogio:IEDM2019}, ReRAM~\cite{Shafiee:ISCA2016}) are used to improve performance per Watt of vector-matrix multiplications by $\sim10^3$ over conventional CMOS ASIC designs. One example of a NVM-based dot-product engine is shown in Figure~\ref{f:dna_dot_product}, where the array consists of word-lines (WLs), bit-lines (BLs) and NVM cells. Each cell on a BL is programmed into a certain resistance ($R$), e.g., $cell_{2x}$ on $BL_2$ is written to $R_{2x}$, where $x=0,1,2$. The cell conductance ($G$) is the inverse of the cell resistance ($\frac{1}{R}$), e.g., $cell_{2x}$ has a conductance of $G_{2x} = \frac{1}{R_{2x}}$. A voltage ($V_x$) can be applied to each WL, so that the current, e.g., $I_{2x}$, passing through a cell ($cell_{2x}$) to the BL is the product of the voltage and the cell conductance ($V_x\cdot G_{2x}$). Based on the Kirchhoff's law, the total current (e.g., $I_2$) on a BL ($BL_2$) is the sum of currents passing through each cell on the BL, so $I_2=\sum_0^2(V_x\cdot G_{2x})$. All BLs in the array produce the current sums simultaneously with the same voltage inputs along WLs. In this way, in each cycle, a vector-matrix multiplication between the input vector $V$ and the conductance matrix $G$ stored in the array is computed by the dot-product engine. The conversion between analog and digital signals is necessary for dot-product engines to communicate with other digital circuits. A digital-analog converter (DAC) converts digital inputs into corresponding voltages that are applied to each WL, while an ADC converts the outputs of a dot-product engine, i.e., the BL accumulated currents, to digital values.

\begin{figure}[h!]
\vspace{-0.1in}
  \centering
	\begin{minipage}{.46\linewidth}
    \centering
		\includegraphics[width=1.5in]{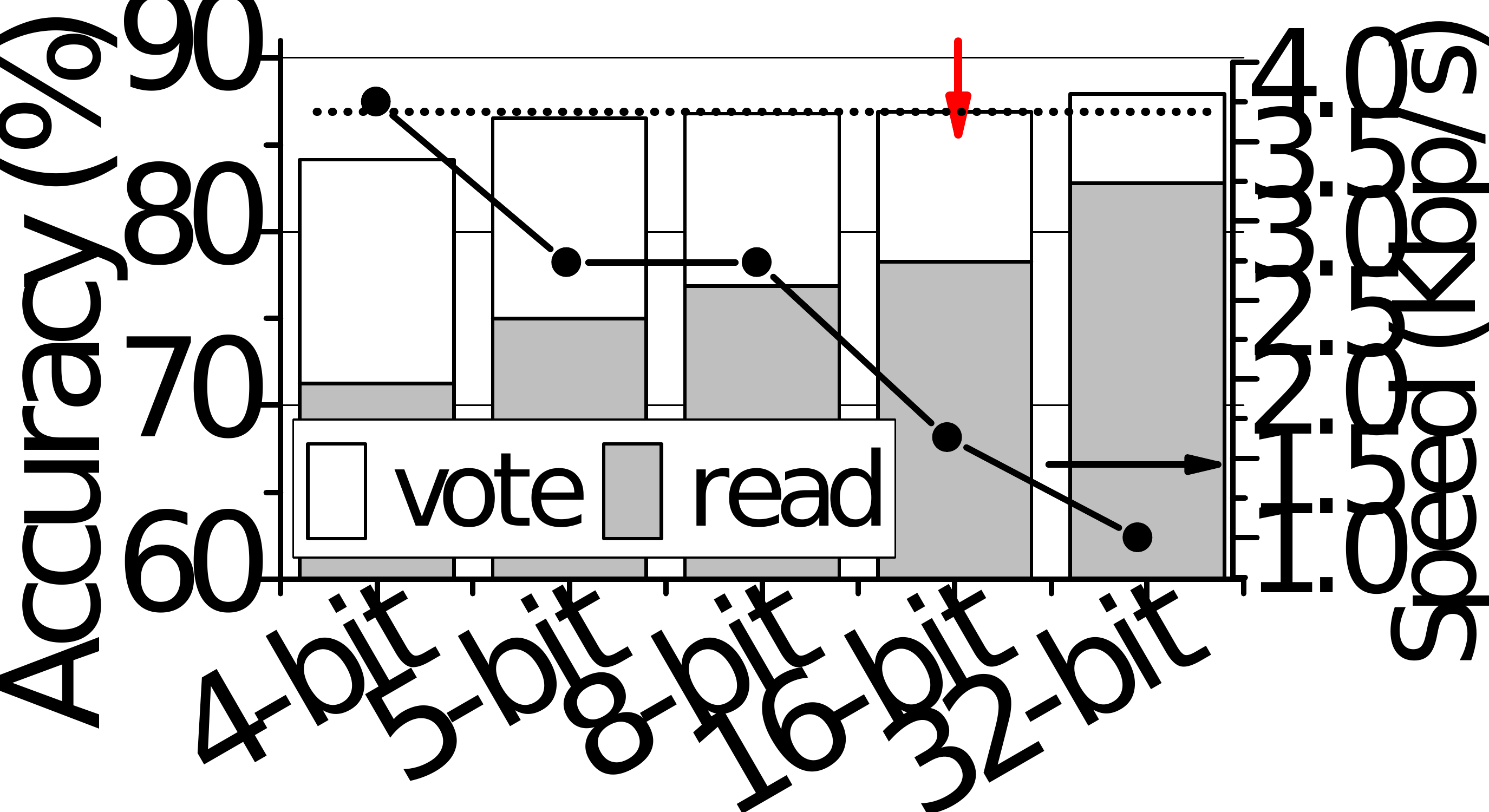}
    \caption{The accuracy \& speed of quantized Guppy.}
		\label{f:dna_quan_me}
  \end{minipage}
	\hspace{0.1in}
  \begin{minipage}{.46\linewidth}
    \centering
    \includegraphics[width=1.5in]{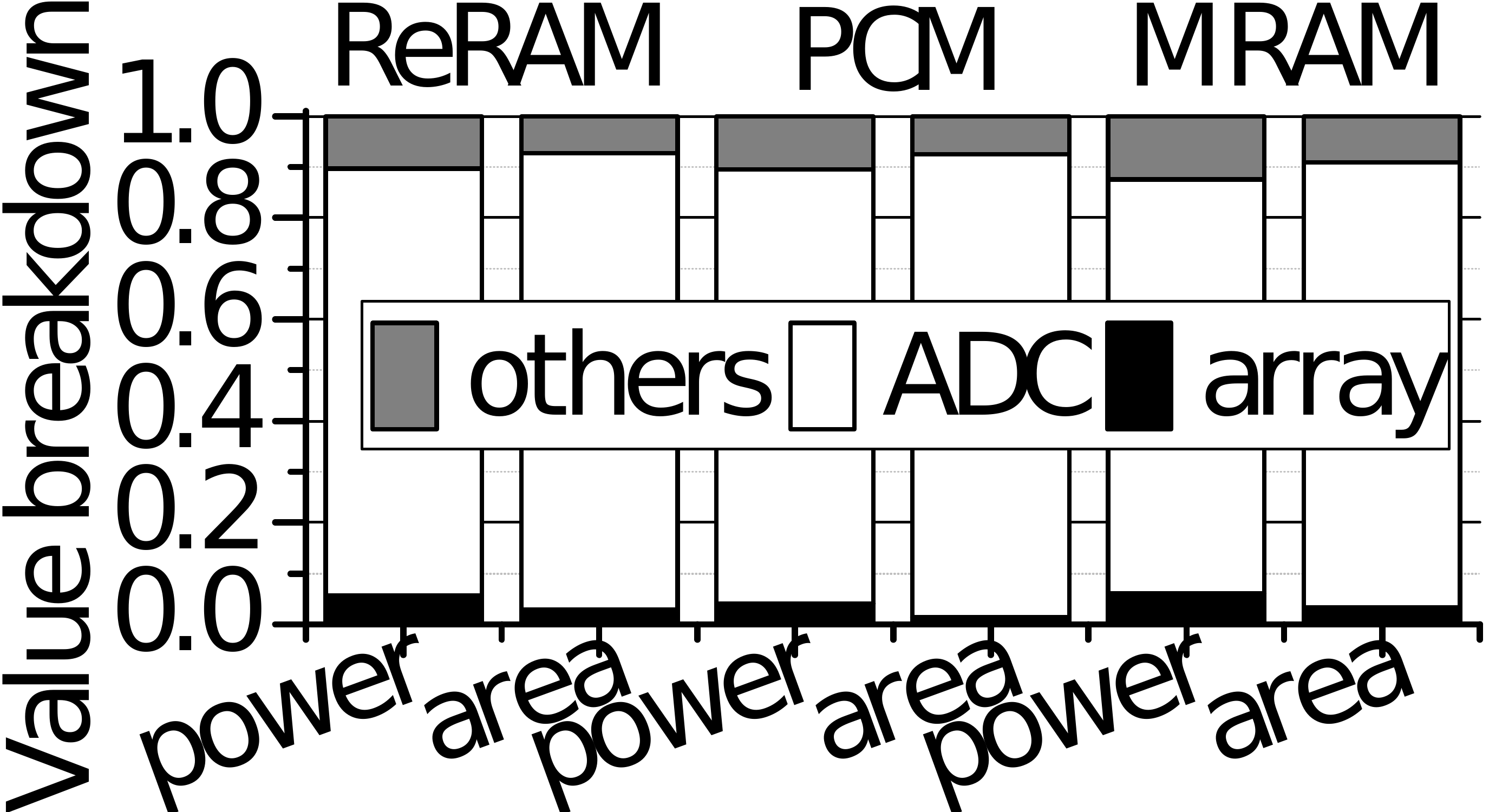}
    \caption{The area and power breakdown of NVM engines.}
		\label{f:dna_pim_break}
  \end{minipage}
\vspace{-0.2in}
\end{figure}

\subsection{SOT-MRAM}

Spin Orbit Torque MRAM (SOT-MRAM)~\cite{Honjo:IEDM2019} emerges as one of the most promising nonvolatile memory alternatives to power hungry SRAM. To record data, SOT-MRAM uses a heavy metal and a perpendicular Magnetic Tunnel Junction (MTJ) consisting of two ferromagnetic layers separated by a thin insulator (MgO), as shown in Figure~\ref{f:stt_sot_cell}. A reference layer has a fixed magnetic direction, while the magnetic direction of the free layer can be switched by an in-plane current flowing through the heavy metal. When two layers have parallel magnetic direction, the MTJ has low resistance state (LRS) and indicates ``0''. In contrast, if two layers are in anti-parallel direction, the MTJ has high resistance state (HRS) and represents ``1''. To write a cell, a write word-line (WWL) is first activated. When the write bit-line (WBL) voltage is larger than the source line (SL) voltage by a threshold, ``1'' is written to the cell. On the contrary, if the WBL voltage is smaller than the SL voltage by a threshold, ``0'' is written to the cell. To read a cell, a read word-line (RWL) is activated, read voltage is applied on the read bit-line (RBL) and the SL is grounded. 

\subsection{Integration of NVM Technologies}

Most emerging NVM technologies, e.g., SOT-MRAM~\cite{Honjo:IEDM2019}, PCM~\cite{Ambrogio:IEDM2019}, ReRAM~\cite{Shafiee:ISCA2016}, are generally CMOS-compatible, so they can be integrated with each other and CMOS logic in the same chip. For instance, a MTJ, i.e., the core of a SOT-MRAM cell, is successfully fabricated with ReRAM cells in a single chip~\cite{Zhang:AEM2018}. Furthermore, the monolithic 3D stacking technology~\cite{Sabry:IEEE2019} can also integrate various NVM technologies including ReRAM and STT-MRAM into a 3D vertical memory array to offer complementary tradeoffs among high density, low latency, and long endurance.

\section{Motivation}
\label{s:motivation}

It is challenging to accelerate nanopore base-calling from both algorithm and architecture perspectives. If we na\"ively accelerate a base-caller using prior network quantization techniques, the quantized base-caller greatly increases the number of systematic errors that cannot be corrected by read voting. State-of-the-art NVM-based PIMs suffer from huge power consumption and area overhead of CMOS ADCs, when executing a quantized base-caller. New bottlenecks, CTC decoding and read voting operations, emerge in a quantized base-caller, but no prior PIM supports these operations.

\begin{figure}[t!]
\centering
\includegraphics[width=3.3in]{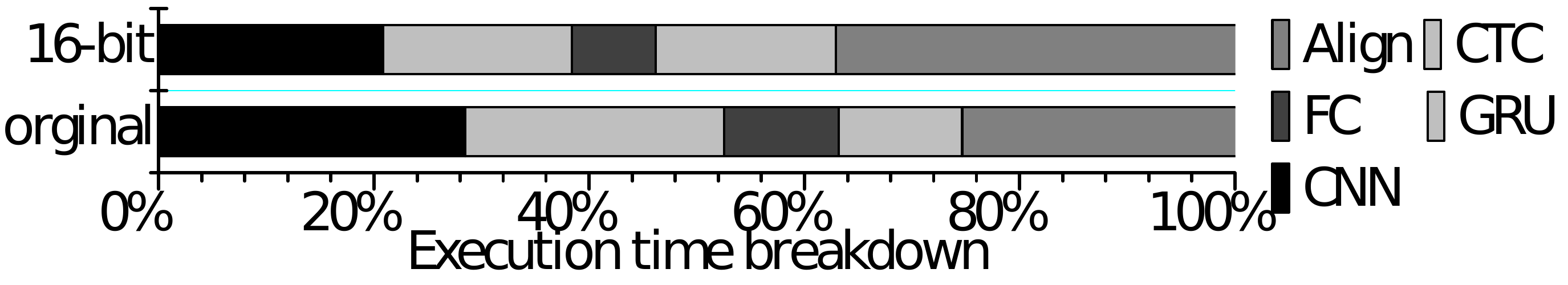}
\vspace{-0.1in}
\caption{Execution time breakdown of Guppy.}
\label{f:dna_time_break}
\vspace{-0.2in}
\end{figure}

\subsection{More Systematic Errors in a Quantized Base-caller}

We applied the latest network quantization technique, FQN~\cite{Li:CVPR2019}, on Guppy to improve its base-calling speed. As Figure~\ref{f:dna_quan_me} shows, the Conv, GRU, FC, and CTC layers of Guppy are quantized with various bit-widths from 4-bit to 32-bit. We executed the quantized Guppy on an NVIDIA Tesla T4 GPU. Although quantizing Guppy with a smaller bit-width, e.g., 4-bit, increases base-calling throughput by $2.75\times$, base-calling accuracy of the quantized Guppy after reads vote decreases by 4.3\%, which dramatically jeopardizes the quality of final DNA mappings. The base-calling accuracy includes two parts: one is the \textit{read accuracy} before reads vote; the other is the \textit{vote accuracy} after reads vote. The base-calling accuracy after reads vote is more important, since read voting operations eliminate all random errors and leave only systematic errors. Even the 16-bit quantized Guppy suffers from significant systematic errors that cannot be corrected by read voting operations.


\begin{figure*}[t!]
\centering
\subfigure[Full-precision model training]{
\includegraphics[width=3.3in]{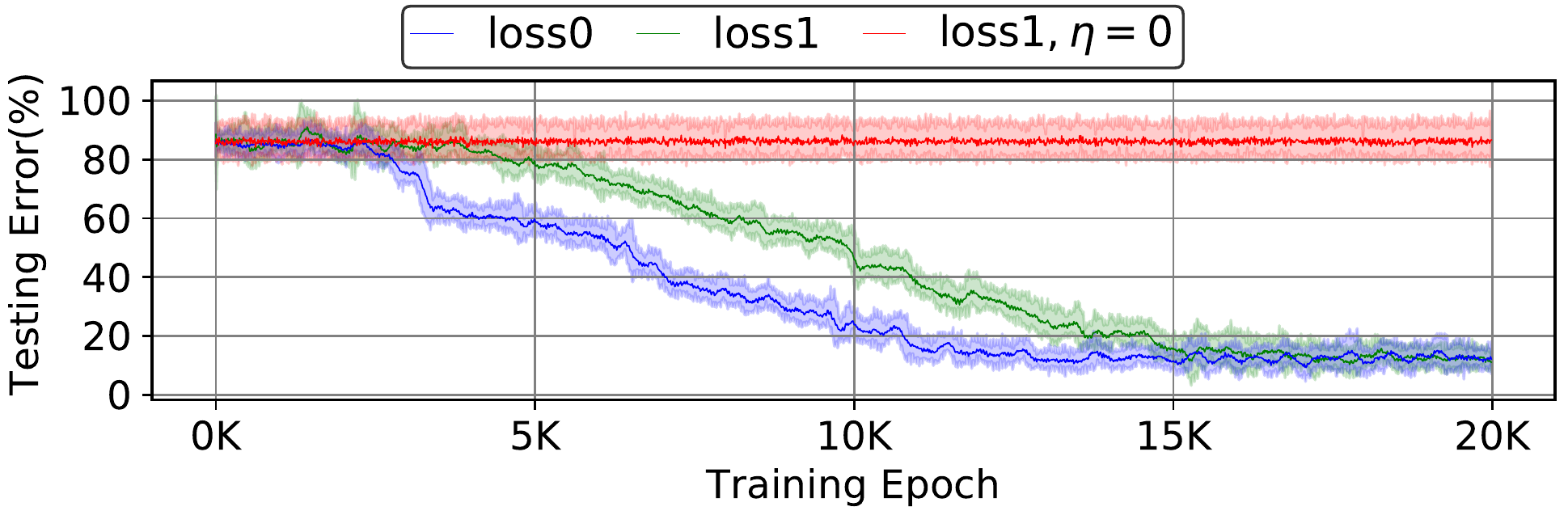}
\label{f:dna_vote_train_full}
}
\subfigure[8-bit quantized model training]{
\includegraphics[width=3.3in]{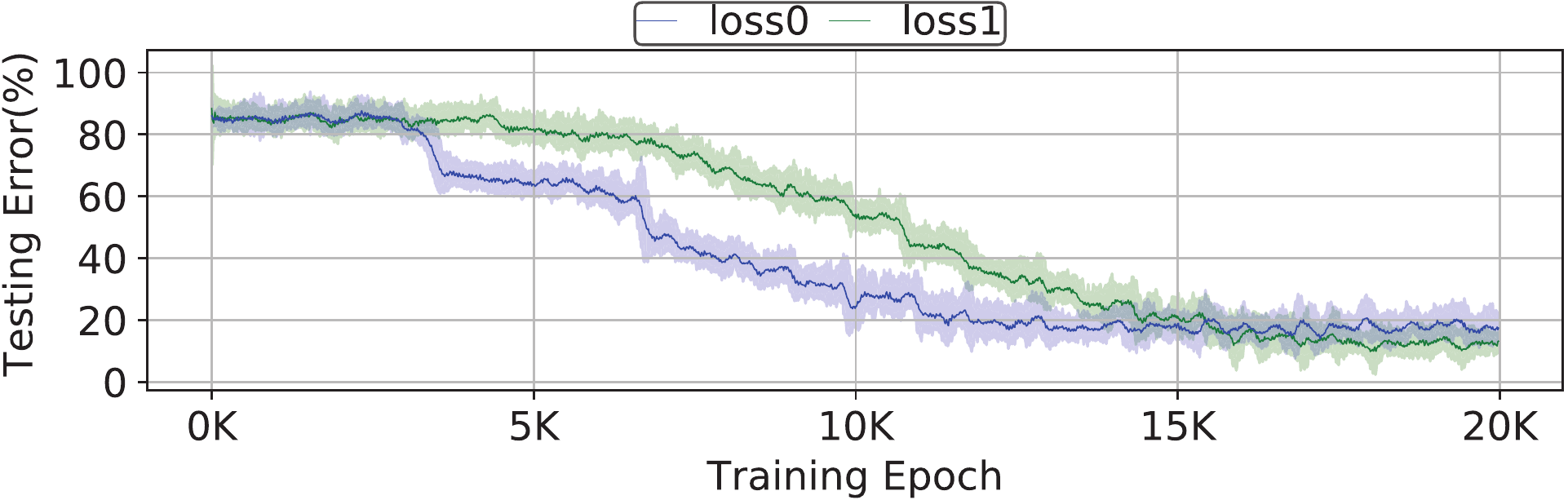}
\label{f:dna_vote_train_8bits}
}
\vspace{-0.1in}
\caption{The training of full-precision and quantized base-callers with different loss functions.}
\label{f:dna_train_effect}
\vspace{-0.2in}
\end{figure*}

\subsection{Large ADC Overhead in NVM-based Dot-product Engines}

Although prior PIM designs process DNN inferences using ReRAM-~\cite{Shafiee:ISCA2016,Fujiki:ASPLOS2018}, PCM-~\cite{Ambrogio:IEDM2019}, and STT-MRAM~\cite{Yan:ICS2018}-based dot-product engines, the power efficiency and scalability of these PIMs are limited by CMOS ADCs. The in-situ analog arithmetic computing fashion is the key for a NVM-based dot-product engine~\cite{Shafiee:ISCA2016,Yan:ICS2018,Ambrogio:IEDM2019,Fujiki:ASPLOS2018} to substantially improving computing throughput of vector-matrix multiplications. However, as Figure~\ref{f:dna_pim_break} highlights, CMOS ADCs cost $82\%\sim85\%$ of power consumption and $87\%\sim91\%$ of area overhead in a ReRAM-~\cite{Shafiee:ISCA2016}, PCM-~\cite{Ambrogio:IEDM2019} and STT-MRAM~\cite{Yan:ICS2018}-based dot-product engine. Although ReRAM, PCM and STT-MRAM has the cell size of $4F^2$, $4F^2$, $60F^2$, respectively, the power and area of array in various NVM dot-product engines are similar, since peripheral circuits including row decoders, column multiplexers and sense amplifiers dominate power consumption and area overhead of a dot-product engine. As a result, CMOS ADCs cost 58\% of power consumption and 30\% of chip area in a typical NVM-based PIM design~\cite{Shafiee:ISCA2016}. The power density of recent NVM-based PIMs has already exceeded the memory thermal tolerance even with active heat sinks. Particularly, a $416W$ ReRAM-based PIM~\cite{Fujiki:ASPLOS2018} has the power density of $842mW/mm^2$, much larger than the thermal tolerance of a ReRAM chip with active heat sinks~\cite{Zhu:MEMSYS2016}. CMOS ADCs seriously limit the scalability and power-efficiency of state-of-the-art NVM-based PIM accelerators.

\begin{figure}[htbp!]
\vspace{-0.1in}
\centering
\includegraphics[width=3.3in]{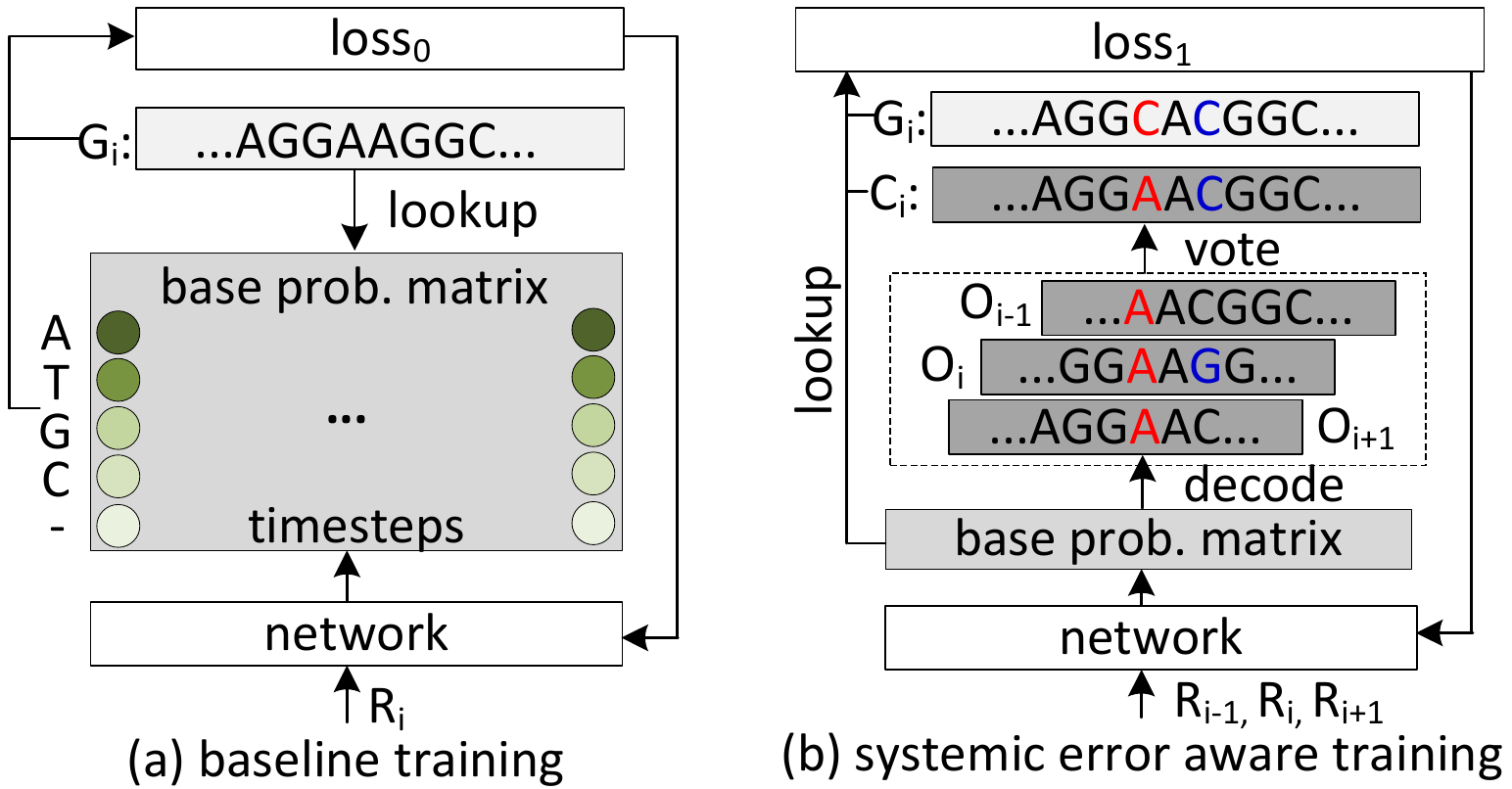}
\vspace{-0.1in}
\caption{Systematic error aware training.}
\label{f:dna_consensus_alg}
\vspace{-0.2in}
\end{figure}

\subsection{New Bottlenecks in a Quantized Base-caller}

Besides more systematic errors, new performance bottlenecks emerge in a 16-bit quantized Guppy. As Figure~\ref{f:dna_time_break} shows, CTC decoding operations consume 16.7\% of base-calling latency, while read voting operations cost 37\% of base-calling latency in the 16-bit quantized Guppy. The Conv, GRU and FC layers in the quantized Guppy heavily rely on 16-bit fixed-point vector-matrix multiplications that can be efficiently executed by a state-of-the-art GPU. Therefore, we anticipate these Conv, GRU and FC layers can be completed by a NVM-based PIM with a shorter latency. In contrast, CTC decoding and read voting operations of a base-caller are not fully optimized on the GPU. Moreover, no prior PIM design supports CTC decoding or read voting.

\section{Helix}
\label{s:brawl}

\subsection{Systematic Error Aware Training}
\label{s:sys_aware}

To reduce the systematic errors that cannot be corrected by read votes, we propose \textit{Systematic Error Aware Training} (SEAT) that aims to minimize the edit distance between a consensus read and its ground truth DNA sequence by a novel loss function during the training of a quantized base-caller.

\textbf{Baseline training}. During the training of a base-caller~\cite{Wick:GB2019,Flappie:ONT2019}, the gradient is not computed through the edit distance between the predicted DNA sequence and its corresponding ground truth, since the computation of edit distance is non-differentiable. As Figure~\ref{f:dna_consensus_alg}a shows, the Conv, GRU and FC layers generate the base probability matrix by an input signal vector $R_i$. Instead of edit distances, the CTC decoder~\cite{Flappie:ONT2019,Teng:GigaScience2018} computes the probability of the ground truth read $G_i$, $p(G_i|R_i)$, as the loss function by applying Equation~\ref{e:dna_prob_cal} on the base probability matrix. For a training set $\mathbb{D}$, the weights of the base-caller are tuned to minimize:
\begin{equation}
loss_0=\sum_{(G_i,R_i)\in\mathbb{D}}(-\ln p(G_i|R_i))
\label{e:dna_train_tar}
\end{equation}
where the more similar to $G_i$ the predicted read is, the smaller $-\ln(p(G_i|R_i))$ is. By making each predicted read more similar to the ground truth, state-of-the-art base-callers indirectly minimizes the number of random and systematic errors. However, random errors can be corrected by read voting operations, whereas only systematic errors are the ``real'' errors that degrade the quality of final DNA mappings.

\textbf{Systematic-error-aware training}. The number of systematic errors significantly increases in a quantized base-caller. We created SEAT for the quantized base-caller to minimize the number of systematic errors. SEAT is shown in Figure~\ref{f:dna_consensus_alg}b. The base-caller uses multiple input signal data vectors, i.e., $R_{i-1}$, $R_i$, and $R_{i+1}$, to generate multiple predicted reads, i.e., $O_{i-1}$, $O_i$, and $O_{i+1}$, that vote to create a consensus read $C_i$. Instead of minimizing the edit distance between $C_i$ and the ground truth read $G_i$, we build a new loss function to make $C_i$ more similar to $G_i$. For a training set $\mathbb{D}$, the parameters of the base-caller are tuned to minimize:
\begin{equation}
\begin{split}
loss_1= \sum_{(G_i,R_i)\in\mathbb{D}} & [-\eta \cdot\ln p(G_i|R_i) + \\
                                      & (\ln p(G_i|R_i)-\ln p(C_i|R_i))^2]
\label{e:dna_train_sys}
\end{split}
\end{equation}
where $-\ln p(G_i|R_i)$ makes each predicted read more similar to $G_i$; $(\ln p(G_i|R_i)-\ln p(C_i|R_i))^2$ minimizes the probability difference between the consensus read $C_i$ voted by multiple predicted reads and $G_i$; and $\eta\in [0, 1]$ is a floating-point constant regulating the impact of $-\ln p(G_i|R_i)$.

\textbf{The effect of SEAT}. As Figure~\ref{f:dna_vote_train_full} shows, we trained a full-precision Guppy by Equation~\ref{e:dna_train_tar} ($loss_0$) and Equation~\ref{e:dna_train_sys} ($loss_1$). If we set $\eta$ in $loss_1$ to 0, the training cannot converge, since it has no motivation to improve the accuracy of each read. When we set $\eta$ to 1, compared to $loss_0$, $loss_1$ slows down training convergence. When the read error rate is high, it is faster to improve the quality of each read independently. However, two loss functions achieve similar base-calling accuracy at the end of the training of Guppy. Full-precision Guppy is powerful enough to minimize the number of systematic errors even without read voting operations. In contrast, the training of 8-bit quantized Guppy with $loss_0$ and $loss_1$ is shown in Figure~\ref{f:dna_vote_train_8bits}. For the 8-bit quantized Guppy, compared to $loss_0$, $loss_1$ increases base-calling accuracy by 6\% and obtains the same base-calling accuracy as the full precision model. After the systematic error reduction capability of Guppy is damaged by network quantization, $loss_1$ can reduce the systematic errors for the quantized Guppy.

\begin{figure}[t!]
\centering
\includegraphics[width=2.5in]{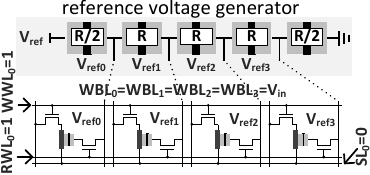}
\vspace{-0.1in}
\caption{The ADC SOT-MRAM array.}
\label{f:dna_dac_array}
\vspace{-0.1in}
\end{figure}

\begin{figure}[t!]
  \centering
  \begin{minipage}{.45\linewidth}
    \centering
    \includegraphics[width=1.65in]{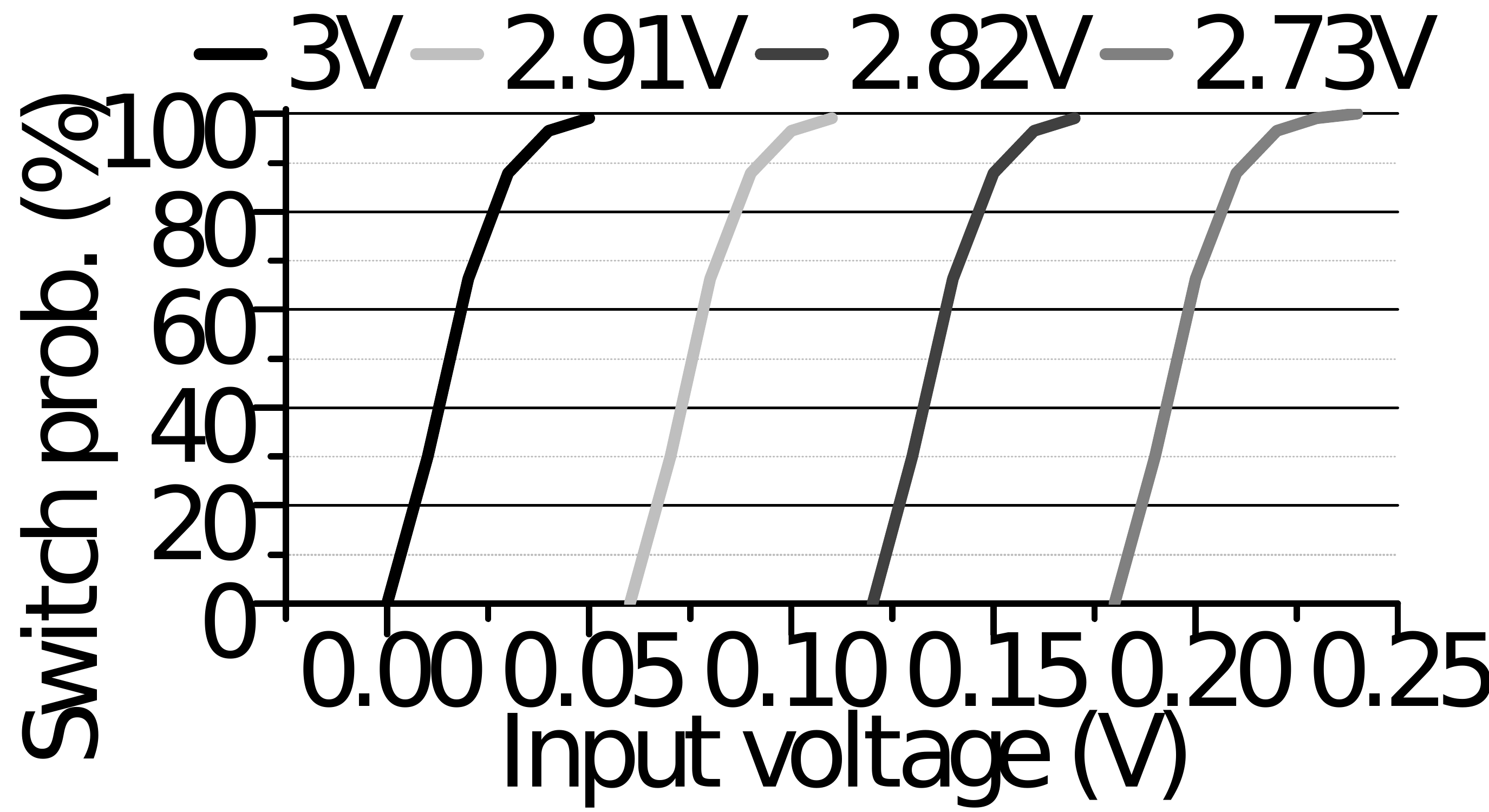}
		\vspace{-0.2in}
    \caption{Input voltage vs. RBL voltage.}
		\label{f:dna_adc_all}
  \end{minipage}\hspace{0.1in}
  \begin{minipage}{.45\linewidth}
    \centering
		\includegraphics[width=1.65in]{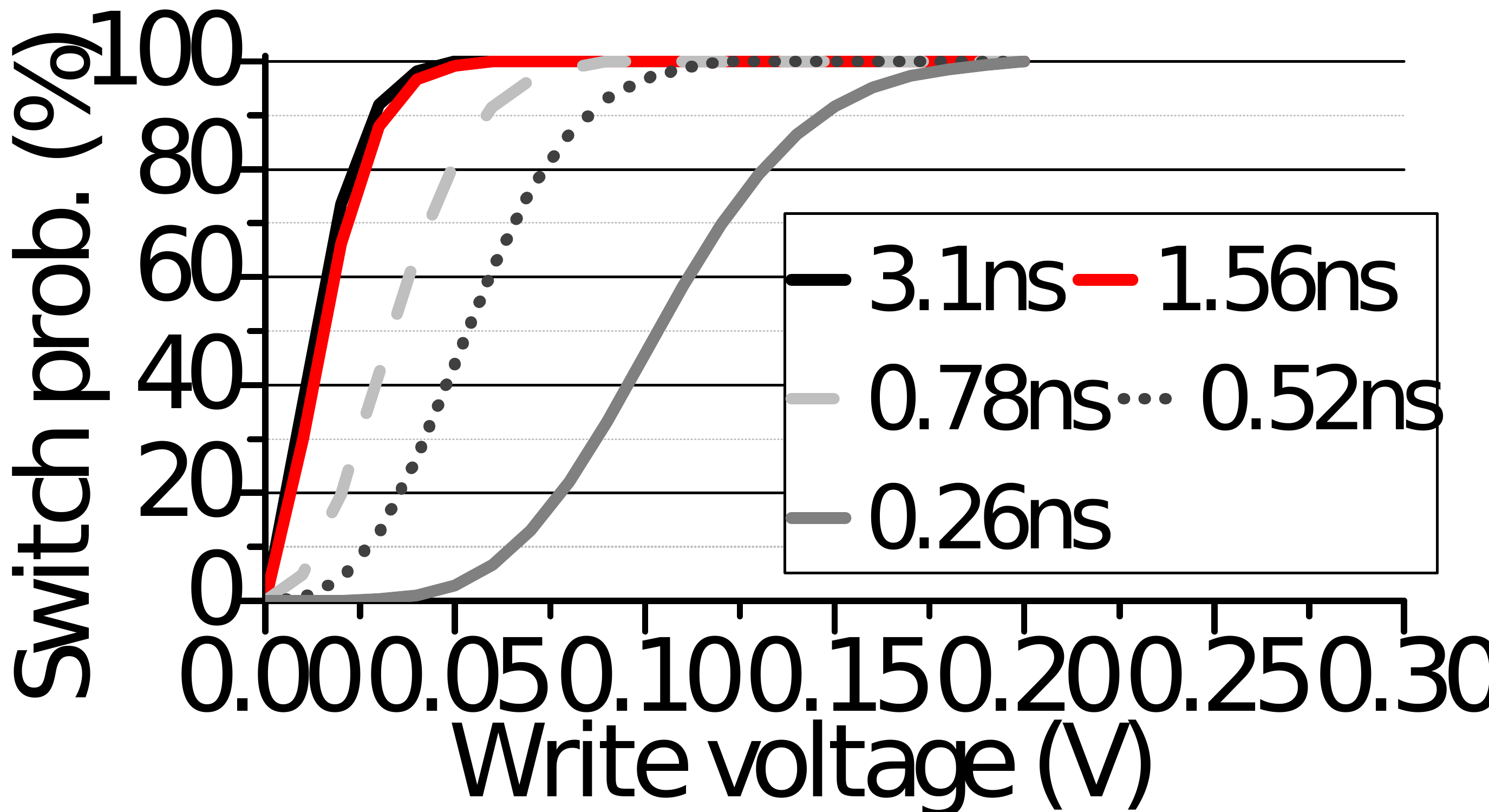}
		\vspace{-0.2in}
    \caption{Write voltage vs. pulse duration.}
		\label{f:dna_adc_write}
  \end{minipage}
\vspace{-0.2in}
\end{figure}

\subsection{ADC-free PIM Accelerator}
\label{s:spin_pim_acc}

To reduce area overhead and power consumption of CMOS ADCs in prior NVM-based PIM accelerators, we propose a SOT-MRAM-based ADC array to reliably process analog-to-digital conversions.

\textbf{ADC array}. An example of a 2-bit ADC array is shown in Figure~\ref{f:dna_dac_array}. To distinguish 2 bits, an ADC array produces four reference voltages ($[V_{ref0} - V_{ref3}]=[3V,2.91V,2.82V,2.73V]$) by a MTJ-based reference voltage generator. In the ADC array, all write word-lines (WWLs) and read word-lines (RWLs) are set to 1, and source lines (SLs) are set to 0. Input voltages are applied to write bit-lines (WBLs), and reference voltages are assigned to read bit-lines (RBLs). As Figure~\ref{f:dna_adc_all} highlights, due to the spin hall effect and voltage-controlled magnetic anisotropy~\cite{Lee:IML2016}, the write voltages of SOT-MRAM are different under various RBL voltages. When a larger voltage is applied on the RBL, the SOT-MRAM write voltage reduces significantly. There are four cases, i.e., 1000, 1100, 1110 and 1111, when an input voltage writes four cells in the ADC Array. By a small encoder, these four cases are encoded to 0, 1, 2 and 3. In this way, the input voltage is converted to a 2-bit digital value. Although a recent work~\cite{Chakraborty:IML2018} leverages the MTJ stochasticity to build an 8-bit ADC by MTJ, the design relies on CMOS counters and registers that introduce large power consumption and area overhead.

\begin{figure}[t!]
  \centering
  \begin{minipage}{.45\linewidth}
    \centering
    \includegraphics[width=1.65in]{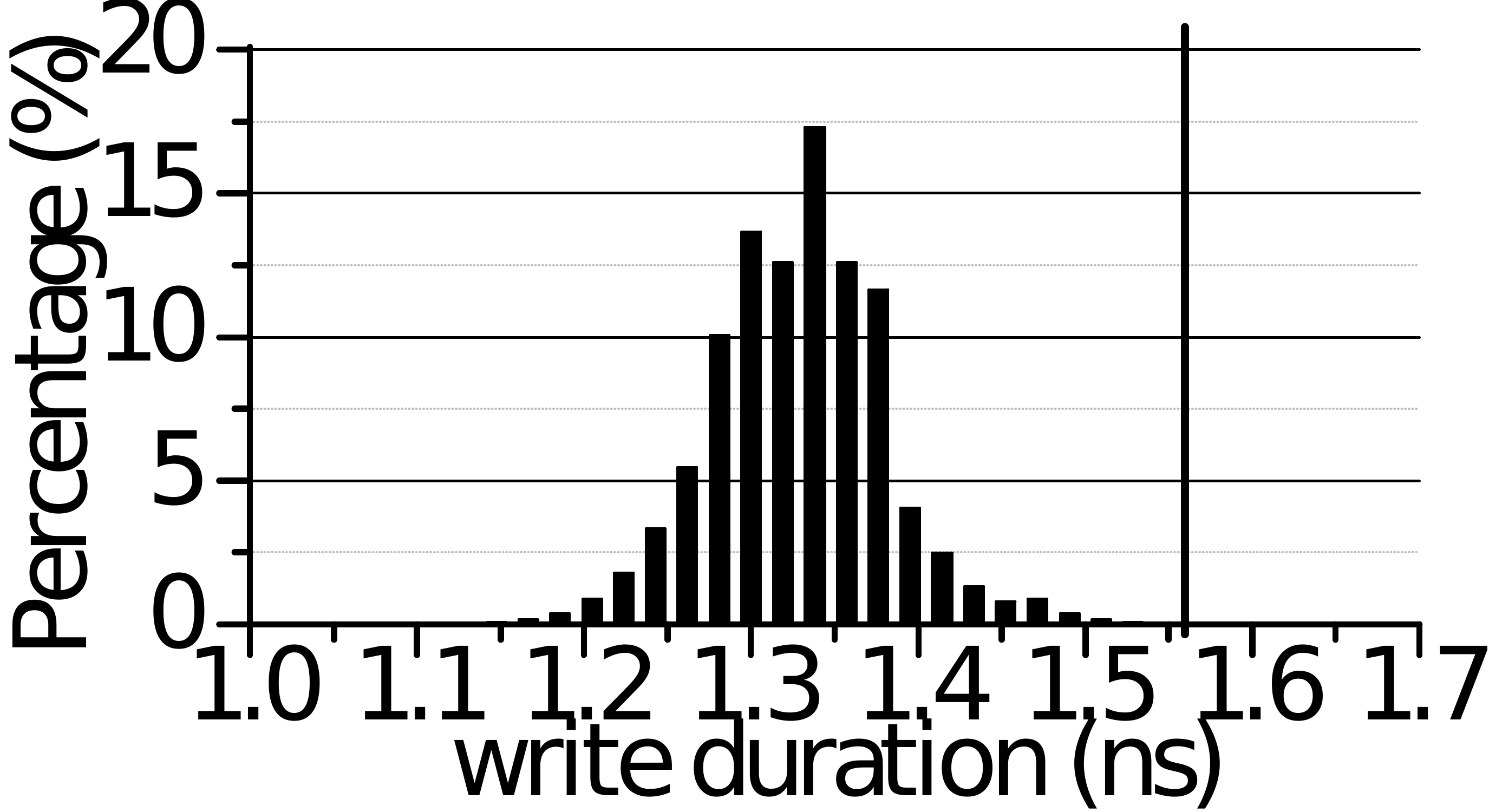}
		\vspace{-0.2in}
    \caption{Write duration with $60F^2$ cell size.}
		\label{f:dna_pv_size}
  \end{minipage}\hspace{0.1in}
  \begin{minipage}{.45\linewidth}
    \centering
		\includegraphics[width=1.65in]{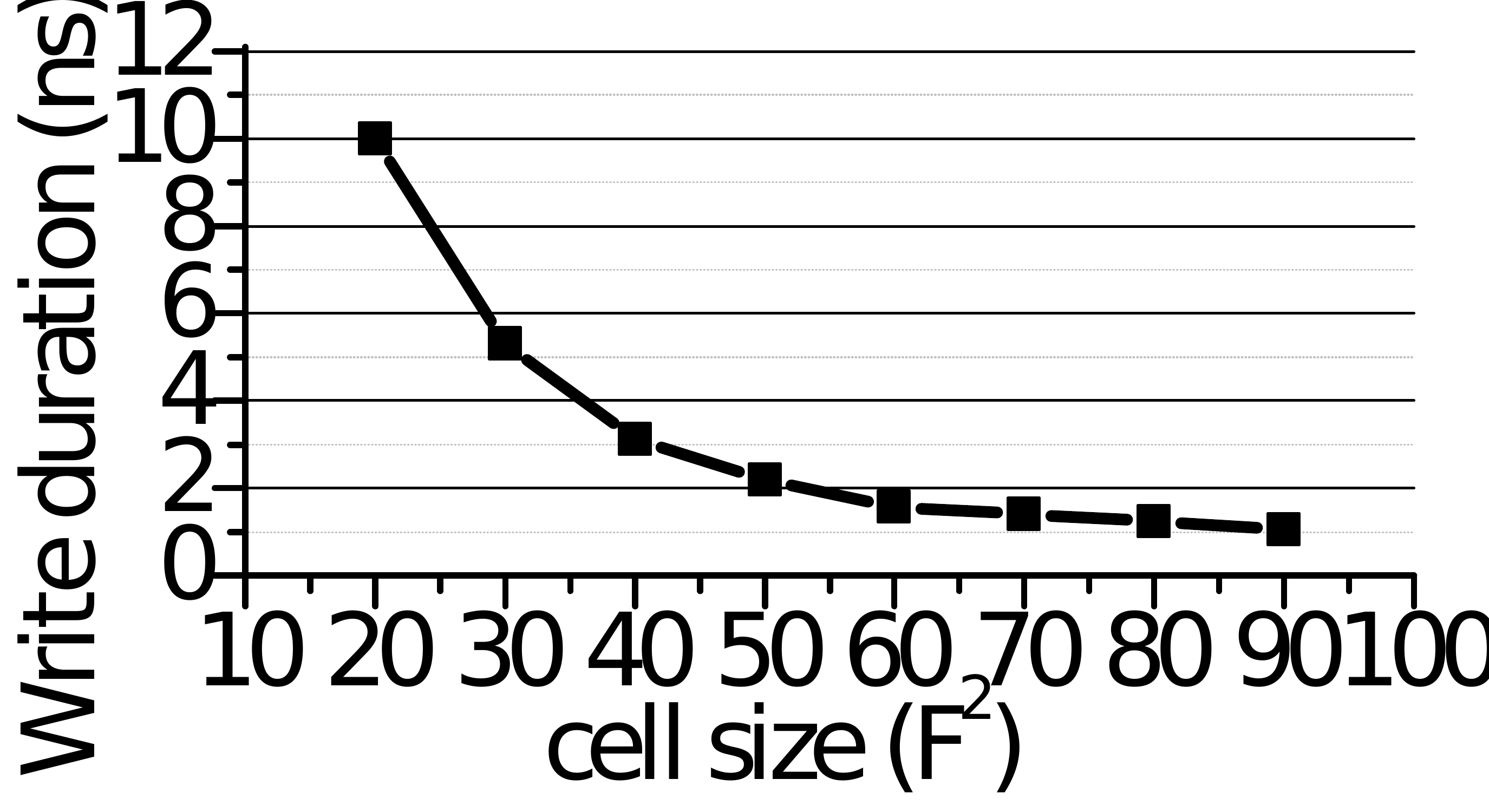}
		\vspace{-0.2in}
    \caption{Worst case duration with varying cell sizes.}
		\label{f:dna_size_late}
  \end{minipage}
\end{figure}

\begin{table}[t!]
\vspace{-0.1in}
\centering
\caption{Process variation of SOT-MRAM}
\vspace{-0.1in}
\setlength{\tabcolsep}{3pt}
\begin{tabular}{|c||c|c|}\hline
Parameter                                & $\mu$                   & $\sigma$        \\\hline\hline
WR/RD transistor width ($W_{wt}$)        & $384nm$                 & $10\%$          \\\hline
WR/RD transistor length ($L_{wt}$)       & $192nm$                 & $10\%$          \\\hline
Threshold voltage ($V_{th}$)             & $0.2V$                  & $10\%$          \\\hline
MTJ resistance area product ($R\cdot A$) & $25\Omega\cdot\mu m^2$  & $8\%$           \\\hline
Cross section area of MTJ ($A$)          & $64nm\times128nm$       & $5\%$           \\\hline
Magnetization stability ($\Delta$)       & $22$                    & $27\%$          \\\hline
\end{tabular}
\label{t:dna_pv_para}
\vspace{-0.1in}
\end{table}

\textbf{Resolution and frequency}. We need to precisely control write pulses in order to enable a higher resolution for the ADC array. There is a trade-off between the resolution and frequency of an ADC array. Figure~\ref{f:dna_adc_write} shows the switching probability of a SOT-MRAM cell under different voltages and pulse durations. The shorter the pulse duration is, the higher frequency an ADC array can be operated at. With a shorter write duration, a higher write voltage is required to reliably switch a cell. Under a fixed maximum input voltage, e.g., $3V$, we can distinguish fewer levels of the input voltage (fewer bits) in Figure~\ref{f:dna_adc_all}. For a higher resolution under $3V$, a smaller write voltage is preferred. In this case, we have to use a longer write pulse duration resulting in lower ADC frequency. To balance the trade-off, we use a $1.56ns$ write pulse to switch a SOT-MRAM cell with $0.05V$. In this way, 32 levels of the input voltage, i.e., 5-bit, can be distinguished. The ADC array can be operated at $640MHz$.

\textbf{Reliability}. SOT-MRAM has no endurance issue, since on average a cell tolerates $10^{15}$ writes~\cite{Kan:ITED2017}. However, process variation makes a SOT-MRAM ADC array to generate wrong outputs. The relation between write current $I$ and pulse duration $t$ can be approximated as
\begin{equation}
t=\tau_0 e^{(1-\frac{I}{A\cdot J_{c0}})\Delta}
\end{equation}
where $A$ is the cross sectional area of the MTJ free layer; $J_{c0}$ is the critical current density at zero temperature; $\Delta$ is the magnetization stability energy height; and $\tau_0$ is a fitting constant. $\Delta$ is decided by the MTJ volume. Due to process variation, different SOT-MRAM cells have different critical parameters including MTJ size, $\Delta$, write transistor width, length and threshold voltage, thereby requiring different write pulse durations. We iteratively increase the write transistor size to guarantee that the worst case cell can be switched in $1.56ns$ by considering process variation. To model the process variation on SOT-MRAM, we adopted the parameters shown in Table~\ref{t:dna_pv_para} from~\cite{Nowak:IML2016}. In each iteration, we conducted 10 billion Monte-Carlo simulations with Cadence Spectre to generate a write duration distribution under a certain SOT-MRAM cell size, which is dominated by the write transistor size. At last, we show the relation between the worst case cell write duration and the cell size in Figure~\ref{f:dna_size_late}. We selected $60F^2$ to tolerate process variation and guarantee the worst case cell write duration is $1.56ns$.

\begin{figure}[t!]
\centering
\includegraphics[width=\linewidth]{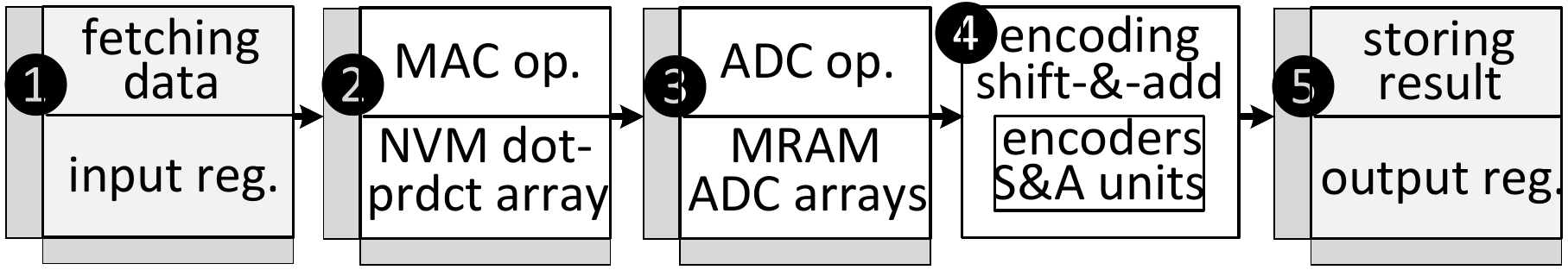}
\vspace{-0.2in}
\caption{The pipeline of a NVM-based dot-product engine.}
\label{f:dna_all_pipeline}
\vspace{-0.2in}
\end{figure}

\textbf{Pipelined dot-product engine}. SOT-MRAM ADC arrays can be easily integrated with prior NVM-based dot-product engines. As Figure~\ref{f:dna_all_pipeline} shows, the pipeline of a fixed-point vector-matrix multiplication includes fetching data, MAC, ADC, shift-\&-add, and storing result. \ding{182} During the stage of fetching data, 128 1-bit fixed-point inputs are read from input registers. The 2-bit weights are stored in a $128\times 128$ array of a NVM-based dot-product engine. \ding{183} A NVM-based dot-product engine converts 1-bit fixed-point inputs to analog voltages by DACs, and performs 1-bit$\times$2-bit matrix-vector multiplications~\cite{Shafiee:ISCA2016}. \ding{184} Multiple ADC arrays digitize a MAC result. The NVM-based dot-product engine generates 128 MAC results simultaneously. \ding{185} After encoding, digital values are sent to shift-\&-add units to generate final dot-product results. \ding{186} At last, the final dot-product results are written into output registers.

\begin{figure}[ht!]
\vspace{-0.1in}
\centering
\includegraphics[width=2.2in]{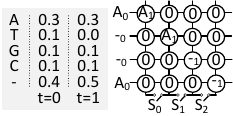}
\vspace{-0.05in}
\caption{CTC decoding in a NVM dot-product engine.}
\label{f:dna_ctc_decoder}
\vspace{-0.2in}
\end{figure}

\subsection{CTC Decoding and Read Vote}
\label{s:dna_read_vote}

\textbf{CTC decoding}. To process CTC beam searches, we rely on a NVM-based dot-product array. Figure~\ref{f:dna_ctc_decoder} shows how to process a CTC beam search with width of 2. The top-2 largest probabilities of bases (i.e., $A_1$ and $-_1$) at the time step 1 ($t=1$) in the CTC base probability matrix are written to the diagonal line cells of a NVM-based dot-product array. Since the search width is 2, each probability of a base at $t=1$ is written twice in two different cells in the diagonal line of the dot-product array. All the other cells in the array are initialized to 0s. We can input the top-2 largest probabilities of bases (i.e., $A_0$ and $-_0$) at $t=2$ to the corresponding WLs, so that $p(A_0A_1)$, $p(A_0-_1)$, $p(-_0A_1)$, and $p(-_0-_1)$ can be concurrently computed. To support the merges of probabilities of multiple-base sequences, we proposed to add a transistor to each BL to connect itself and its neighboring BL. By closing all transistors ($S_0\sim S_2$), we merged the probabilities of four 2-base sequences. In this way, we have $p(A)=p(A_0A_1)+p(A_0-_1)+p(-_0A_1)+p(-_0-_1)$. 


\textbf{Reliability of NVM dot-product arrays}. Since each BL has only one base's probability, the resistance of the transistor we add on each BL is too small to introduce errors in CTC decoding. Since a NVM dot-product array can operate at only 10MHz~\cite{Shafiee:ISCA2016}, the extra transistor does not slow down the dot-product array. However, our design increases writes to a NVM dot-product array. A ReRAM cell stands for $10^{11}$ writes. A recent ReRAM-based PIM~\cite{Fujiki:ASPLOS2018} can reliably run back-propagation for 15.7 years. Compared to back-propagation, the Conv, GRU, FC layers and a CTC decoder of a base-caller have much less writes. Based on our estimation, the NVM dot-product arrays of Helix can reliably work for $>$20 years even when running Chiron having the most complex architecture and the largest number of parameters among all base-callers.

\begin{figure}[t!]
\centering
\includegraphics[width=3.1in]{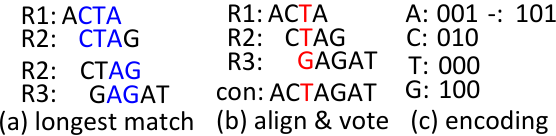}
\vspace{-0.1in}
\caption{Read voting.}
\label{f:dna_vote_read}
\vspace{-0.2in}
\end{figure}

\textbf{Read vote}. After a base-caller generates multiple consecutively predicted reads, a read vote is required to produce a consensus read. A voting example is shown in Figure~\ref{f:dna_vote_read}, where there are three reads, i.e., $R_1$=``ACTA'', $R_2$=``CTAG'', and $R_3$=``GAGAT''. A vote finds the longest matches between all reads (Figure~\ref{f:dna_vote_read}a), aligns reads, and computes the consensus (Figure~\ref{f:dna_vote_read}b). Finding the longest matches between all reads is the most important operation in a read vote. To find the longest match between $R_1$ and $R_2$, all of their sub-strings have to be compared. As Figure~\ref{f:dna_vote_read}(c) describes, we encoded each DNA symbol by 3-bit. The string match problem is converted to comparing two binary vectors. We propose a SOT-MRAM-based binary comparator array to accelerate binary vector comparisons.

\begin{figure}[ht!]
\vspace{-0.1in}
\centering
\includegraphics[width=3.3in]{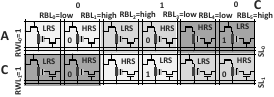}
\vspace{-0.1in}
\caption{A binary comparator array.}
\label{f:dna_xnor_sam}
\vspace{-0.1in}
\end{figure}

\textbf{Binary comparator array}. We wrote all sub-strings of $R_1$, e.g. ``ACTA'' and ``CTA'', into a SOT-MRAM array shown in Figure~\ref{f:dna_xnor_sam}. Each sub-string stays in a row of the array. For instance, ``ACTA'' is in the first row, while ``CTA'' is in the second row. We used a 2-cell pair in a row to record each bit in the encoding of a DNA symbol. 0 is represented by a low resistance state (LRS) cell and a high resistance state (HRS) cell, while 1 is indicated by a HRS cell and a LRS cell. Therefore, in Figure~\ref{f:dna_xnor_sam}, 6 cells in the first row indicate the first ``A'' of ``ACTA'', while 6 cells in the second row represent the first ``C'' of ``CTA''. We applied the corresponding voltages representing a sub-string of $R_2$, e.g., ``C'', on the RBLs of the binary comparator array. Each bit in the encoding of ``C'' (010) is represented by two voltages applied on the two RBLs of a 2-cell pair respectively, i.e., 0 is represented by low and high voltages, while 1 is denoted by high and low voltages. If two DNA symbols are the same, there is no current accumulated on the SL, e.g., $SL_1$. The sense amplifier can sense a current on the SL, e.g., $SL_0$, if two DNA symbols are different. Unlike alignment and assembly, aligning reads during read voting is easy~\cite{Teng:GigaScience2018}, because the order of these reads is already known and the length of each read is only $10\sim30$ bases.

\textbf{Reliability of binary comparator arrays}. To compare two 30-base reads, a binary comparator array requires $>180$ cells on a RWL. We used the $60F^2$ cell size to build $256\times256$ arrays as binary comparators to study process variation. We also adopted the same process variation parameters in Table~\ref{t:dna_pv_para}. We performed 10 billion Monte-Carlo simulations to profile the error rate with random 30-base read inputs. The error rate for reading a single cell is low, i.e., $10^{-11}$. After comparing 556 million 30-base reads, on average, our binary comparator array makes 1 mistakes. We believe this error rate is acceptable for Helix, since assembly, read mapping, and polishing in the nanopore sequencing pipeline may correct systematic errors.  

\begin{table}[t!]
\centering
\caption{The area and power of Helix}
\vspace{-0.1in}
\setlength{\tabcolsep}{1pt}
\begin{tabular}{|l|l|l|l|l|}
\hline
Component                & Params      & Spec           & Power ($mW$)   & Area ($mm^2$)          \\ \hline\hline

eDRAM                    & bank num    & 4              & 20.7           & 0.083                  \\ 
Buffer                   & capacity    & 64KB           &                &                        \\ \hline
Bus                      & wire num    & 384            & 7              & 0.09                   \\ \hline
Router                   & flit size   & 32             & 10.5           & 0.0378                \\ \hline
Activation               & number      & 2              & 0.52           & 0.0006                 \\
S+A                      & number      & 1              & 0.05           & 0.00006                \\
MaxPool                  & number      & 1              & 0.4            & 0.0024                 \\
OR                       & size        & 3KB            & 1.68           & 0.0032                 \\ \hline
\textbf{Total}           &             &                & 40.9           & 0.215                  \\ \hline\hline

NVM                      & number      & 8              &                &                        \\ 
Array                    & size        & 128$\times$128 & 2.4            & 0.0002                 \\ 
                         & bits/cell   & 2              &                &                        \\ \hline
S+H                      & number      & 8$\times$128   & 0.001          & 0.00004                \\ \hline
S+A                      & number      & 4              & 0.2            & 0.00024                \\ 
IR                       & size        & 2KB            & 1.24           & 0.0021                 \\
OR                       & size        & 256B           & 0.23           & 0.00077                \\ \hline
DAC                      & resolution  & 1 bit          & 4              & 0.00017                \\
                         & number      & 8$\times$128   &                &                        \\ \hline\hline
                         & resolution  & 8 bits         &                &                        \\
ADC                      & frequency   & 1.28 GSps      & 16             & 0.0096                 \\
										     & number      & 8              &                &                        \\ \hline
\textbf{ISAAC Total}     & number      & 12             & 289            & 0.157                  \\ \hline 
\textbf{ISAAC Tile Total}&             &                & 330            & 0.372                  \\ \hline 
\textbf{ISAAC Total}     & tile num    & 168            & \textbf{55.4W} & \textbf{62.5}          \\ \hline\hline

SOT-MRAM                 & size        & 32$\times$32   &                &                        \\ 
ADC array                & frequency   & 640MHz         & 0.6            & 0.00005                \\
                         & number      & 8$\times$4     &                &                        \\ \hline
voltage ref              & number      & 1              & 0.02           & 0.00003                \\
encoder                  & number      & 8$\times$4     & 0.001          & 0.000002               \\ \hline
\textbf{Helix Total}     & number      & 12             & 122            & 0.0439                 \\ \hline
\textbf{Helix Tile Total}&             &                & 163            & 0.259                  \\ \hline 
SOT-MRAM                 & size        & 256$\times$256 & 1.3W           & 0.11                   \\
binary cmp               & number      & 1024           &                &                        \\ \hline
\textbf{Helix Total}     & tile num    & 168            & \textbf{25.7W} & \textbf{43.83}         \\ \hline

\end{tabular}
\label{t:dna_power_area}
\vspace{-0.2in}
\end{table}

\subsection{Design Overhead}
\label{s:overhead}

For the algorithm modification, our systematic error aware training increased the training time of quantized base-callers by $32\%\sim52\%$ ($\sim2$ days). For the NVM PIM design, we developed Helix based on a well-known ReRAM PIM ISAAC~\cite{Shafiee:ISCA2016}, because we showed that PCM-, STT-, and ReRAM-based dot-product engines have similar power consumption and area overhead (Figure~\ref{f:dna_pim_break}). Although recent search efforts on NVM PIMs propose compilation support~\cite{Fujiki:ASPLOS2018}, data flow optimization~\cite{Ankit:ASPLOS2019}, and sparsity reduction~\cite{Yang:ISCA2019}, all their architectures are built upon ISAAC~\cite{Shafiee:ISCA2016}. To estimate the hardware overhead of Helix, we modeled the leakage power, dynamic energy, latency and area of Helix by NVSim~\cite{Dong:TCAD2012} with 32nm process technology. The power consumption and area overhead of Helix is described in Table~\ref{t:dna_power_area}. The NVM dot-product pipeline is operated at 10MHz~\cite{Shafiee:ISCA2016}. 8-bit~\cite{Shafiee:ISCA2016}, 6-bit~\cite{Yang:ISCA2019}, and 5-bit~\cite{Fujiki:ASPLOS2018} ADCs are adopted by prior PIMs. Although we selected 8-bit ADCs in our baseline, we perform a sensitivity study on the ADC resolution in~\cref{s:eval_analy}. To support CTC decoding, we add a transistor to each BL of a NVM-based dot-product engine introducing insignificant power and area overhead. To accelerate read votes, we also integrated 1K 256$\times$256 SOT-MRAM arrays that cost only 1.3W power and occupy 0.11$mm^2$.

\begin{table}[t!]
\centering
\caption{The architecture of various base-callers}
\vspace{-0.1in}
\begin{tabular}{|l|l|c|c|c|}
\hline
\multicolumn{2}{|c|}{}                         & Scrappie      & Chiron                       & Guppy   \\ \hline\hline
\multicolumn{2}{|c|}{Input}   & \multicolumn{3}{c|}{$300\times 1$}                                        \\ \hline
\multirow{7}{*}{Conv}         & layer \#       & 1             & 3                            & 1                          \\ \cline{2-5}
                              & filter size    & $11\times1$   & $1\times1/3$                 & $11\times1$                \\ \cline{2-5}
                              & filter \#      & 96            & 256                          & 96                         \\ \cline{2-5}
						                  & strides        & 5             & 1                            & 2                          \\ \cline{2-5}
					                    & output         & $60\times96$  & $60\times256$                & $150\times96$              \\ \cline{2-5}
															& MAC \#         & 0.063M        & 570M                         & 0.2736M                    \\ \cline{2-5}
					                    & Param \#       & 1056          & 1.9M                         & 0.0018M                    \\ \hline
															
\multirow{6}{*}{RNN }         & type           & GRU           & LSTM                         & GRU                        \\ \cline{2-5}
                              & layer \#       & 5             & 6                            & 5                          \\ \cline{2-5}
                              & filter         & 96            & 100                          & 256                        \\ \cline{2-5}
					                    & output         & $60\times1025$& $300\times100$               & $150\times40$              \\ \cline{2-5}
															& MAC \#         & 8.1M          & 45M                          & 36M                        \\ \cline{2-5}
					                    & Param \#       & 0.14M         & 0.15M                        & 0.23M                      \\ \hline

\multirow{5}{*}{FC}           & layer \#       & 1             & 1                            & 1                          \\ \cline{2-5}
                              & filter         & $1025\times5$ & $100\times5$                 & $40\times5$                \\ \cline{2-5}
                              & output         & $60\times5$   & $300\times5$                 & $60\times5$                \\ \cline{2-5}
															& MAC \#         & 0.31M         & 0.15M                        & 0.012M                        \\ \cline{2-5}
					                    & Param \#       & 0.31M         & 0.15M                        & 0.012M                       \\ \hline
		
\multicolumn{2}{|c|}{CTC}            & \multicolumn{3}{c|}{output $60\times 1$ and then merge}\\ \hline
\multicolumn{2}{|c|}{Align}          & \multicolumn{3}{c|}{align multiple reads}\\ \hline
\multicolumn{2}{|c|}{Total MAC \#}   & 8.47M        & 615.2M     & 36.3M                   \\ \hline
\multicolumn{2}{|c|}{Total Param \#} & 0.45M        & 2.2M       & 0.244M                    \\\hline
\end{tabular}
\label{t:dna_basecall_arch}
\vspace{-0.2in}
\end{table}

\section{Experimental Methodology}
\label{s:exp_meth}

\subsection{Simulation and Evaluation}

We adopted a NVM dot-product engine simulator from~\cite{Yang:ISCA2019} and modified it to cycle-accurately study the performance, power and energy consumption of Helix and our baseline NVM-based PIM accelerator. According to a user-defined accelerator configuration and a DNN topology description, the simulator generates the performance and power details of the accelerator inferring the DNN. We integrated the ADC array and binary comparator arrays of Helix into the pipeline and data flow of the simulator. We implemented our systematic error aware training in base-callers~\cite{Teng:GigaScience2018,Wick:GB2019,Scrappie:ONT2019} that are trained on either an NVIDIA Tesla T4 GPU or an Intel Xeon E5-4655 v4 CPU.

\subsection{Base-callers and Datasets}
\label{s:dna_other_callers}

\textbf{Base-callers}. Oxford nanopore technology had updated its pore type to R9.4. Among all base-callers, only Metrichor~\cite{Metrichor:Online2018}, Albacore~\cite{Oxford:Albacore2018}, Flappie~\cite{Flappie:ONT2019}, Scrappie~\cite{Scrappie:ONT2019}, Guppy~\cite{Wick:GB2019}, and Chiron~\cite{Teng:GigaScience2018} can base-call R9.4 reads. Metrichor is a cloud-based base-caller whose details are unknown, while Albacore is deprecated by Oxford nanopore technology. Albacore has been replaced by its GPU-version successor Guppy and CPU-version successor Flappie. Guppy and Flappie share the same DNN topology. In this paper, we include three base-callers: Guppy, Scrappie, and Chiron. Guppy and Chiron are GPU-based base-callers, while Scrappie can be executed on only a CPU. We redesigned Scrappie using TensorFlow, so that it can also be processed by a GPU. The base-caller architectures can be viewed in Table~\ref{t:dna_basecall_arch}. All base-callers share a similar network architecture including convolutional, recurrent neural network (RNN), and fully-connected layers. The RNN can be a GRU or Long Short Term Memory (LSTM) layer. Chiron has the most complex DNN topology. Particularly, its convolutional layers have the largest number of weights, while its RNN is a LSTM layer having more recurrent gates. We assume the beam search width of the CTC decoder in each base-caller is 10.

\begin{table}[htbp!]
\centering
\caption{The dataset for various base-callers.}
\vspace{-0.05in}
\setlength{\tabcolsep}{3pt}
\centering
\begin{tabular}{|l||c|c|}
\hline
Sample          & \# of reads & Median read length\\\hline\hline
Phage Lambda    & 34,383      & 5,720 bases\\\hline
E.coli          & 15,012      & 5,836 bases\\\hline
M.tuberculosis  & 147,594     & 3,423 bases\\\hline
Human           & 10,000      & 6,154 bases\\\hline
\end{tabular}
\label{t:dna_testing_data}
\end{table}

\textbf{Datasets}. We used R9.4 training datasets~\cite{Teng:Chiron2018} including \textit{E. coli}, \textit{Phage Lambda}, \textit{M. tuberculosis} and \textit{human} to train base-callers. The input signal is normalized by subtracting the mean of the entire read and dividing by the standard deviation. At the beginning of each training epoch, the dataset was shuffled first and then fed into the base-caller by batch. Training with this mixed dataset enabled each base-caller to have better performance both on generality and base-calling accuracy. The datasets for the evaluation of various base-callers are summarized in Table~\ref{t:dna_testing_data}.

\begin{table}[htbp!]
\centering
\caption{The comparison between CPU, GPU and Helix.}
\vspace{-0.05in}
\setlength{\tabcolsep}{3pt}
\centering
\begin{tabular}{|l||c|c|c|}
\hline
Parameter       & CPU         & GPU           & Helix     \\\hline\hline
core \#         & 8           & 2560          & 16128     \\\hline
Frequency       & 3.2GHz      & 1.5GHz        & 10MHz     \\\hline
Area            & 450$mm^2$   & 515$mm^2$     & 43.83$mm^2$\\\hline
TPD             & 135W        & 70W           & 25.7W       \\\hline
Cache           & 30MB L3     & 6MB L2        & -         \\\hline
Memory          & 32GB DDR4   & 16GB GDDR6    & 32GB NVDIMM \\\hline    
\end{tabular}
\label{t:dna_baseline_all}
\vspace{-0.2in}
\end{table}

\subsection{Schemes}
We compared our Helix PIM against the state-of-the-art CPU, GPU and NVM PIM baselines summarized as:
\begin{itemize}[noitemsep,topsep=0pt,leftmargin=*]
\item \texttt{CPU}. Our CPU baseline is a 3.2GHz Intel Xeon E5-4655 v4 CPU, which has 8 cores and 30MB last level cache. More details can be viewed in Table~\ref{t:dna_baseline_all}.

\item \texttt{GPU}. We selected NVIDIA Tesla T4 GPU as our GPU baseline, since it can support INT8 and INT4 MAC operations. A 1.5GHz NVIDIA Tesla T4 GPU has 2560 cudaCores and a 16GB GDDR6 main memory.

\item \texttt{ISAAC}. We also chose ISAAC~\cite{Shafiee:ISCA2016} as our NVM PIM baseline. We assumed ISAAC has the same processing throughput of CTC decoding and read vote without introducing extra power consumption and area overhead. By studying the sensitivity of the ADC resolution, we compared Helix against two successors of ISAAC including IMP~\cite{Fujiki:ASPLOS2018} and SRE~\cite{Yang:ISCA2019}.

\item \texttt{16-bit}. We quantized base-callers with 16-bit and without systematic error aware training (SEAT) to achieve no obvious accuracy degradation. The quantized base-callers are ran on \texttt{ISAAC}.

\item \texttt{SEAT}. We quantized base-callers with 5-bit and SEAT to guarantee no accuracy loss. The quantized base-callers are ran on \texttt{ISAAC}.

\item \texttt{ADC}. We replaced CMOS ADCs of \texttt{SEAT} by our proposed ADC arrays.

\item \texttt{CTC}. We used NVM-based dot-product arrays to process CTC decoding operations for \texttt{ADC}.

\item \texttt{Helix}. We used SOT-MRAM-based binary comparator arrays to accelerate read votes for \texttt{CTC}. All techniques we proposed in this paper are accumulated in this scheme.
\end{itemize}

\begin{figure}[ht!]
\vspace{-0.1in}
  \centering
  \begin{minipage}{.45\linewidth}
    \centering
    \includegraphics[width=1.65in]{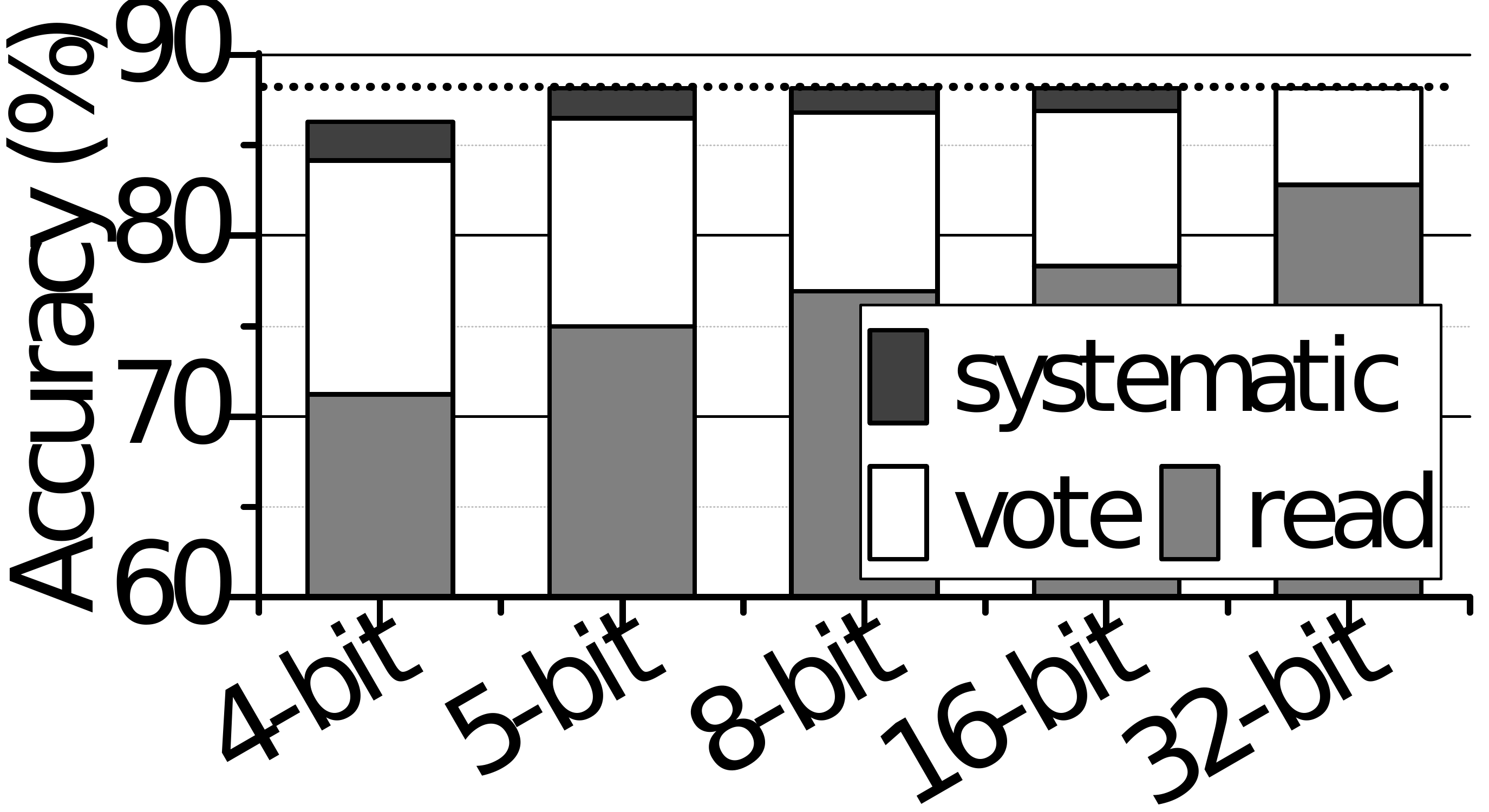}
		\vspace{-0.2in}
    \caption{SEAT on Guppy.}
		\label{f:dna_quan_you}
  \end{minipage}\hspace{0.1in}
  \begin{minipage}{.45\linewidth}
    \centering
		\includegraphics[width=1.65in]{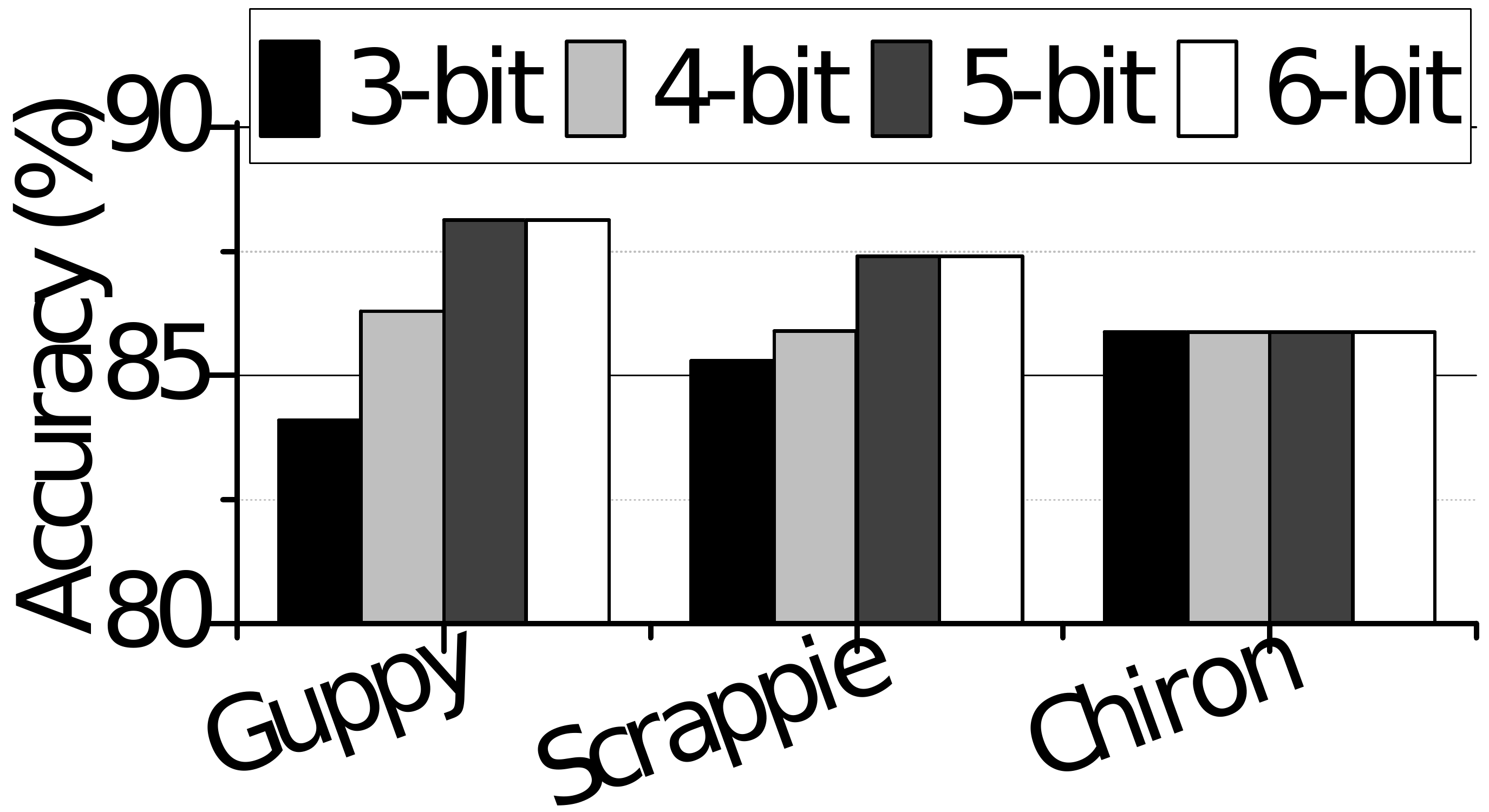}
		\vspace{-0.2in}
    \caption{Quant. w. SEAT.}
		\label{f:dna_base_callers}
  \end{minipage}
\vspace{-0.2in}
\end{figure}

\section{Evaluation and Analysis}
\label{s:eval_analy}

\subsection{Systematic Error Aware Training}

\textbf{SEAT \& quantization}. Though na\"ively applying the quantization scheme FQN~\cite{Li:CVPR2019} on base-callers improves base-calling throughput, the number of systematic errors that cannot be corrected by read votes greatly increases. After we trained Guppy with our systematic error aware training (SEAT), we can reduce the number of systematic errors. As Figure~\ref{f:dna_quan_you} shows, SEAT makes the quantized \texttt{Guppy} have no accuracy loss by reducing the number of systematic errors in its loss function, if it is quantized with $\geq5$-bit. In contrast, without SEAT, the 16-bit quantized \texttt{Guppy} starts to suffer from a significant number of systematic errors. In this way, SEAT enables more aggressive quantization with smaller bit-widths. We show base-calling accuracy of various quantized base-callers in Figure~\ref{f:dna_base_callers}. We find that with 5-bit, no quantized base-caller suffers from accuracy degradation. However, with smaller bit-widths, e.g., 4-bit, \texttt{Scrappie} and \texttt{Guppy} suffer from obvious accuracy degradation, since they have compact architectures and less parameters. The parameter-rich Chiron does not decrease its base-calling accuracy, even when quantized with 3-bit.

\begin{figure}[ht!]
\vspace{-0.1in}
\centering
\includegraphics[width=\linewidth]{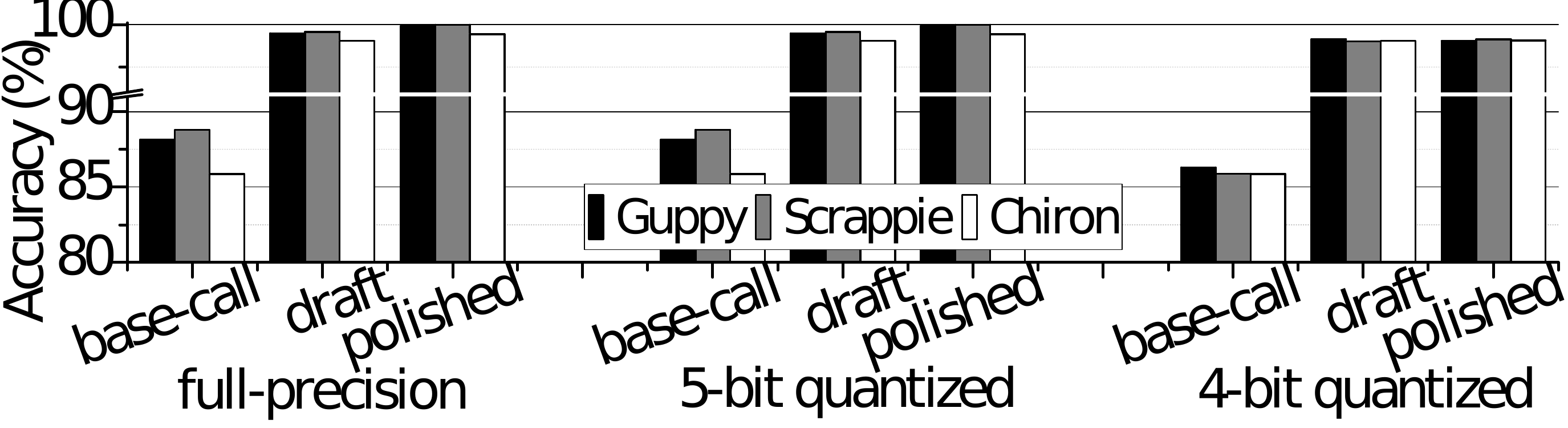}
\vspace{-0.2in}
\caption{The comparison of base-callers with SEAT.}
\label{f:dna_final_acc}
\vspace{-0.1in}
\end{figure}

\begin{figure*}[t!]
\centering
\subfigure[Performance]{
\includegraphics[width=2.2in]{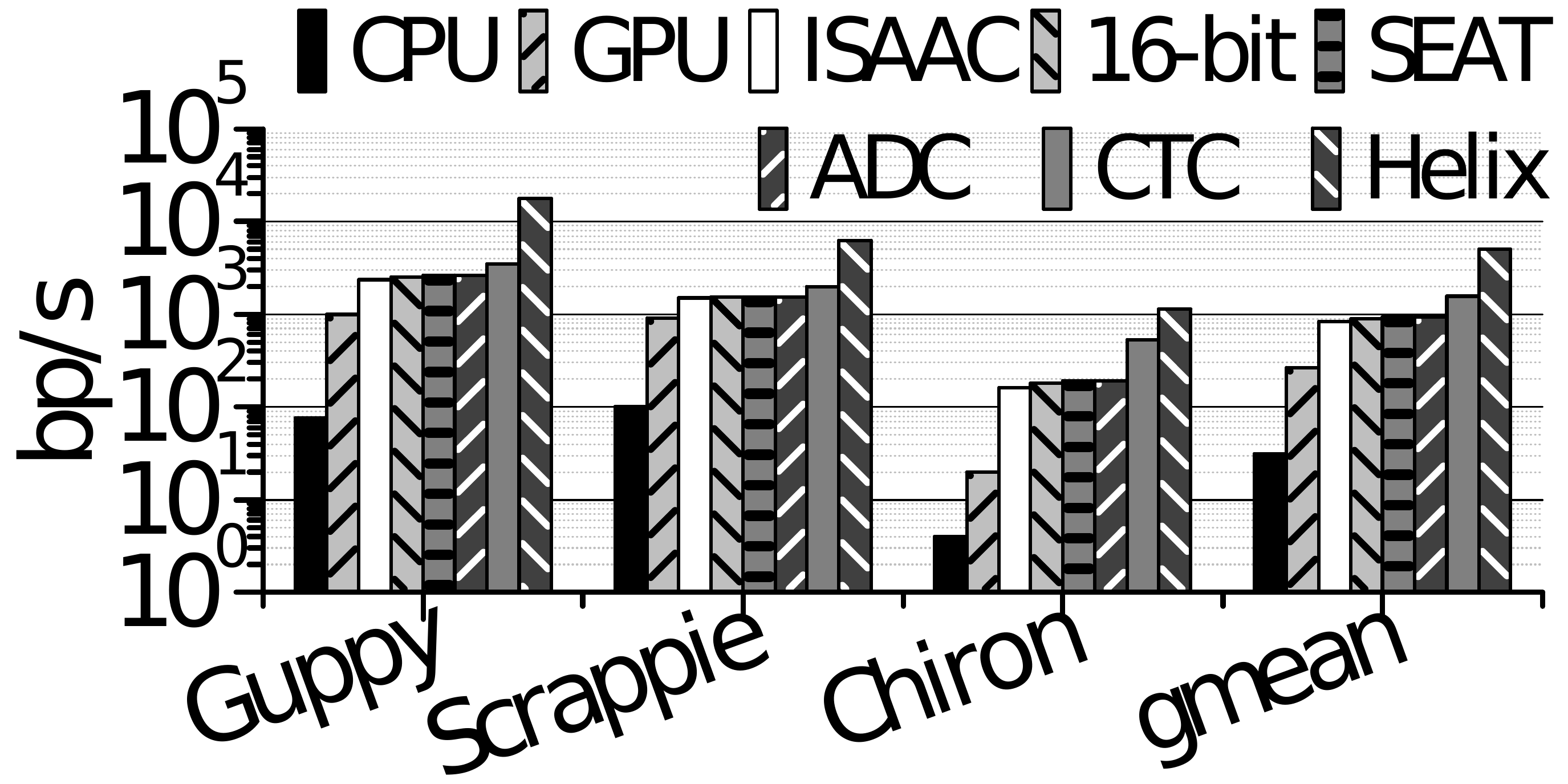}
\label{f:dna_perf_all}
}
\subfigure[Performance/Watt]{
\includegraphics[width=2.2in]{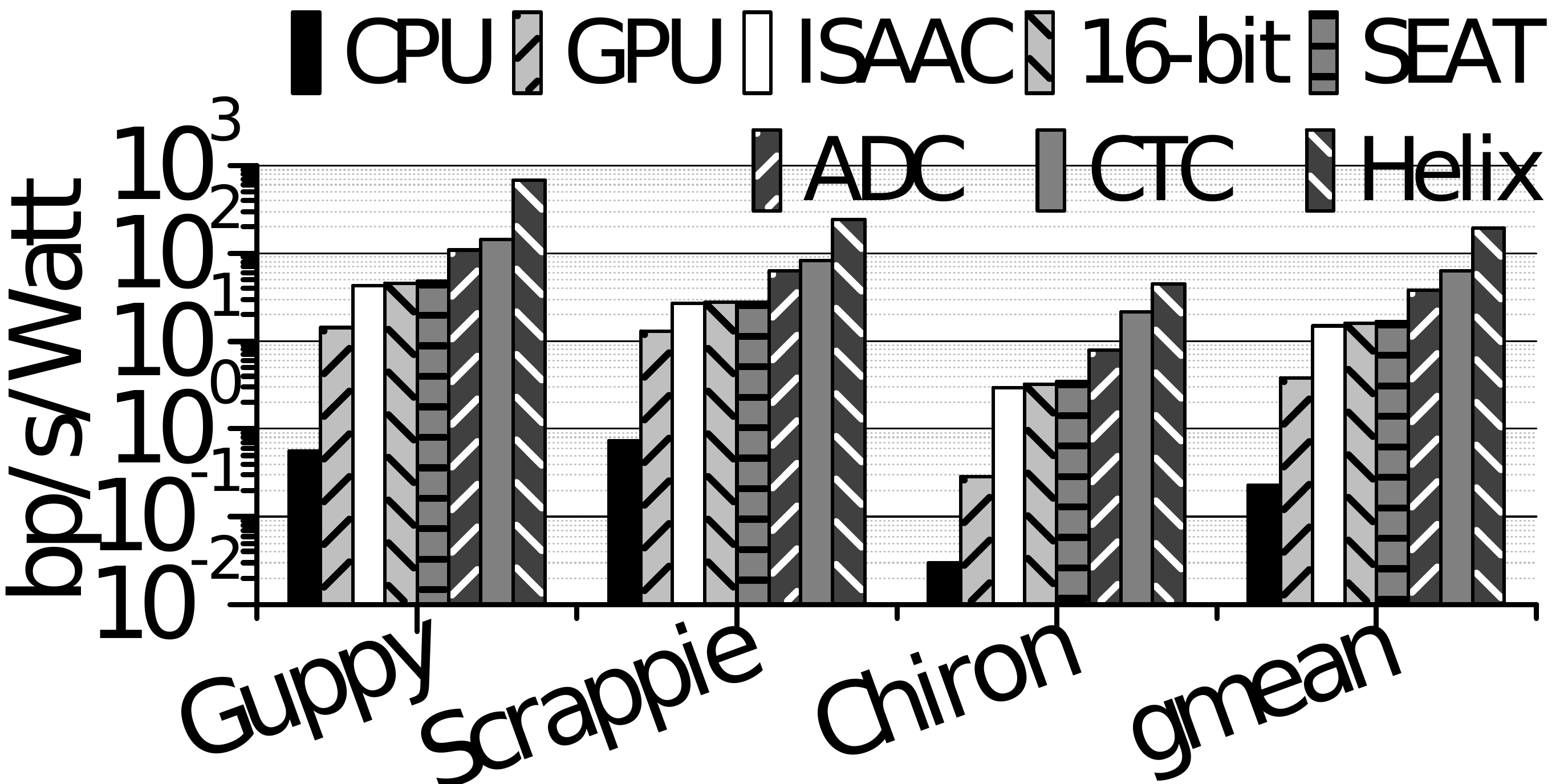}
\label{f:dna_perfwatt_all}
}
\subfigure[Performance/$mm^2$]{
\includegraphics[width=2.2in]{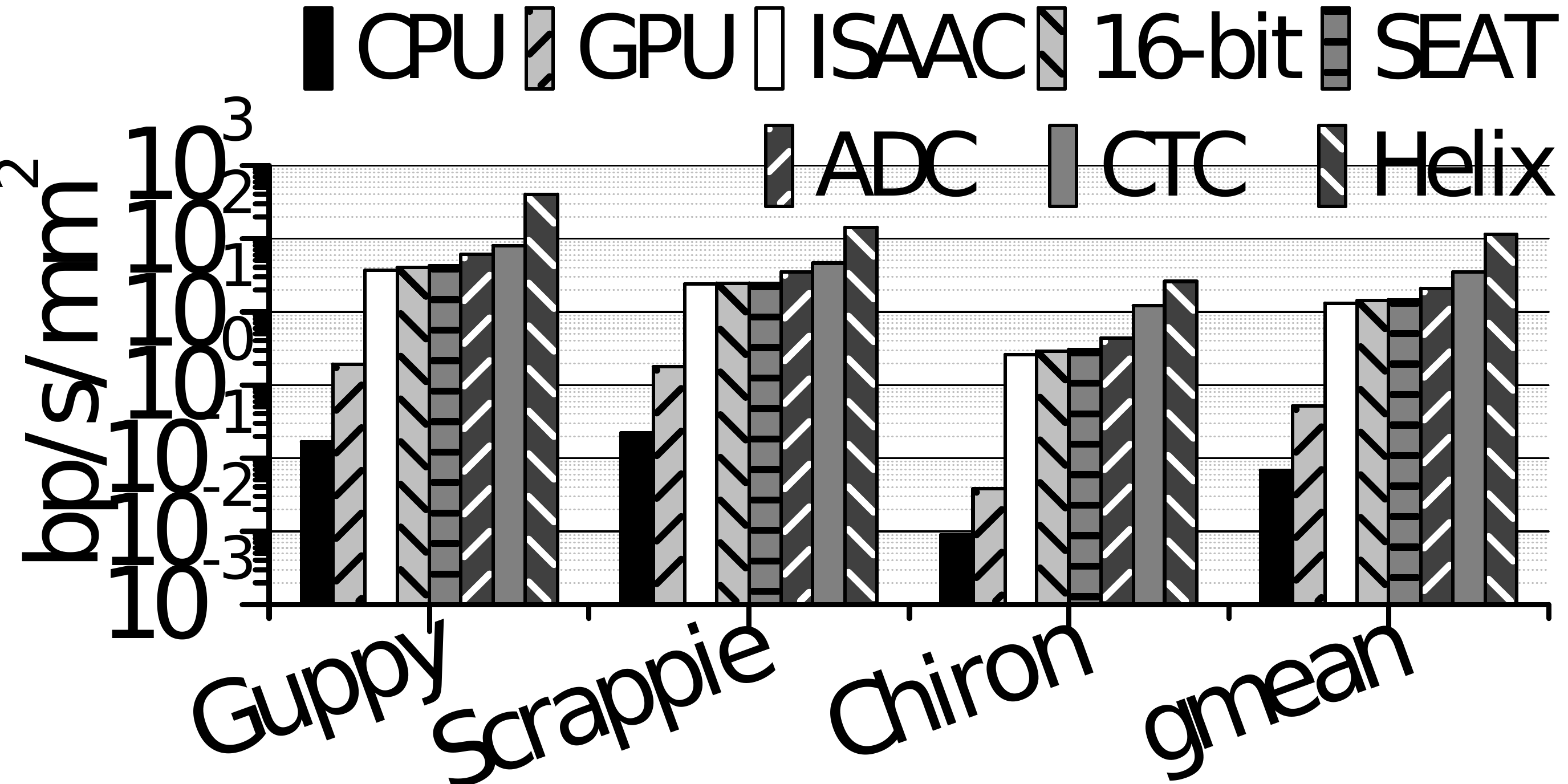}
\label{f:dna_mm_all}
}
\vspace{-0.15in}
\caption{The performance, power and area comparison between various accelerators.}
\label{f:dna_all_allall}
\vspace{-0.15in}
\end{figure*}

\textbf{Quality of final genome mappings}. We fed base-called DNA reads generated by quantized base-callers with SEAT into the nanopore sequencing pipeline to evaluate the quality of final DNA mappings. The accuracy comparison of various DNA mappings generated by both the full-precision, 4-bit, and 5-bit quantized base-callers with SEAT is shown in Figure~\ref{f:dna_final_acc}, where ``base-call'' indicates the accuracy of reads generated by base-callers; ``draft'' represents the accuracy of alignment produced by read mapping; and ``polished'' means the accuracy of final read mapping after the polishing step. Compared to full-precision base-callers, the accuracy of reads, corresponding draft alignment, and final mapping generated by 5-bit quantized base-callers with SEAT has no accuracy loss. However, if we quantize the base-callers with 4-bit, the accuracy of base-called reads, their alignment and final mapping significantly degrades even with SEAT. Particularly, the 4-bit quantized \texttt{Scrappie} reduces the accuracy of the final mapping by 6\%. Low quality genome mappings substantially increased the probability of misdiagnosis and false negative testings. Therefore, we used 5-bit to quantize these base-callers with SEAT.

\textbf{Performance, power and area}. The performance, power and area comparison between our CPU, GPU, and NVM-based PIM baselines is shown in Figure~\ref{f:dna_all_allall}. Besides the \texttt{CPU} and \texttt{GPU}, we ran the DNN part of full-precision base-callers with 32-bit weights on our PIM baseline \texttt{ISAAC}, but left the other parts of base-callers including CTC decoding and aligning on the GPU without introducing extra power consumption and area overhead. As Figure~\ref{f:dna_perf_all} shows, on average, \texttt{ISAAC} greatly improves base-calling throughput by $25\times$ and $2.15\times$ over the CPU and GPU, respectively. Among all base-callers, Chiron achieves the largest speedup by running its DNN part on \texttt{ISAAC}, since 95\% of the base-calling time is consumed by its DNN part. \texttt{ISAAC} improves base-calling throughput of Chiron by $7.16\times$ over \texttt{GPU}. \texttt{ISAAC} also increases base-calling throughput per Watt and per $mm^2$ by $127\%$ and $25\times$ over \texttt{GPU} respectively, as shown in Figure~\ref{f:dna_perfwatt_all} and~\ref{f:dna_mm_all}. If we quantize base-callers with 16-bit, \texttt{16-bit} improves base-calling speed by 6.25\% over \texttt{ISAAC}. On the contrary, if we use SEAT to aggressively quantize base-callers with 5-bit, \texttt{SEAT} improves base-calling speed by 11.1\% over \texttt{ISAAC} without accuracy loss. Although the base-calling throughput improvement achieved by SEAT is not dramatically significant, SEAT is the key to enabling our power-efficient SOT-MRAM ADC arrays with lower resolution.

\begin{figure}[ht!]
\centering
\subfigure[Performance/Watt]{
\includegraphics[width=1.55in]{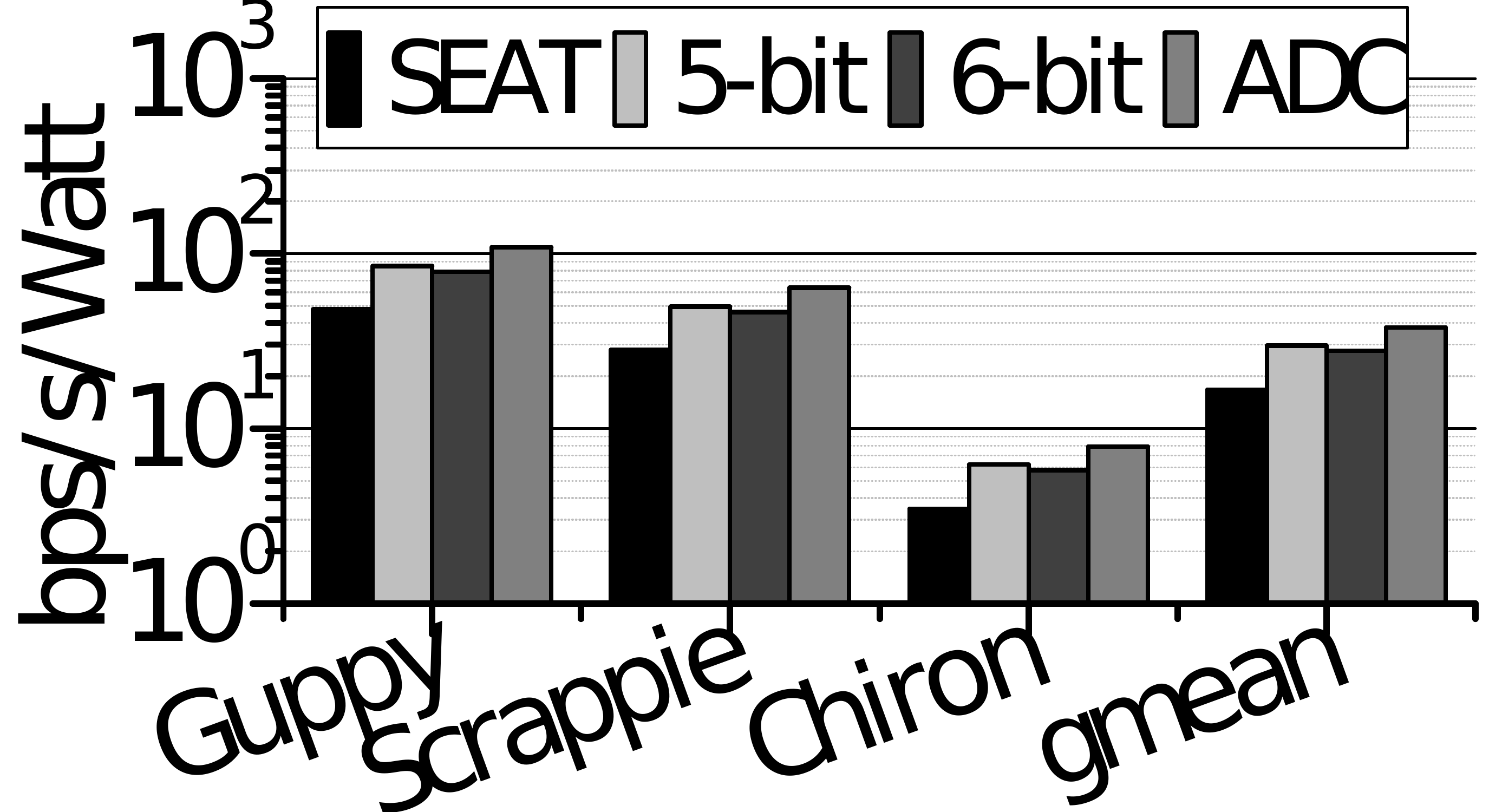}
\label{f:dna_adc_perf}
}
\subfigure[Performance/$mm^2$]{
\includegraphics[width=1.55in]{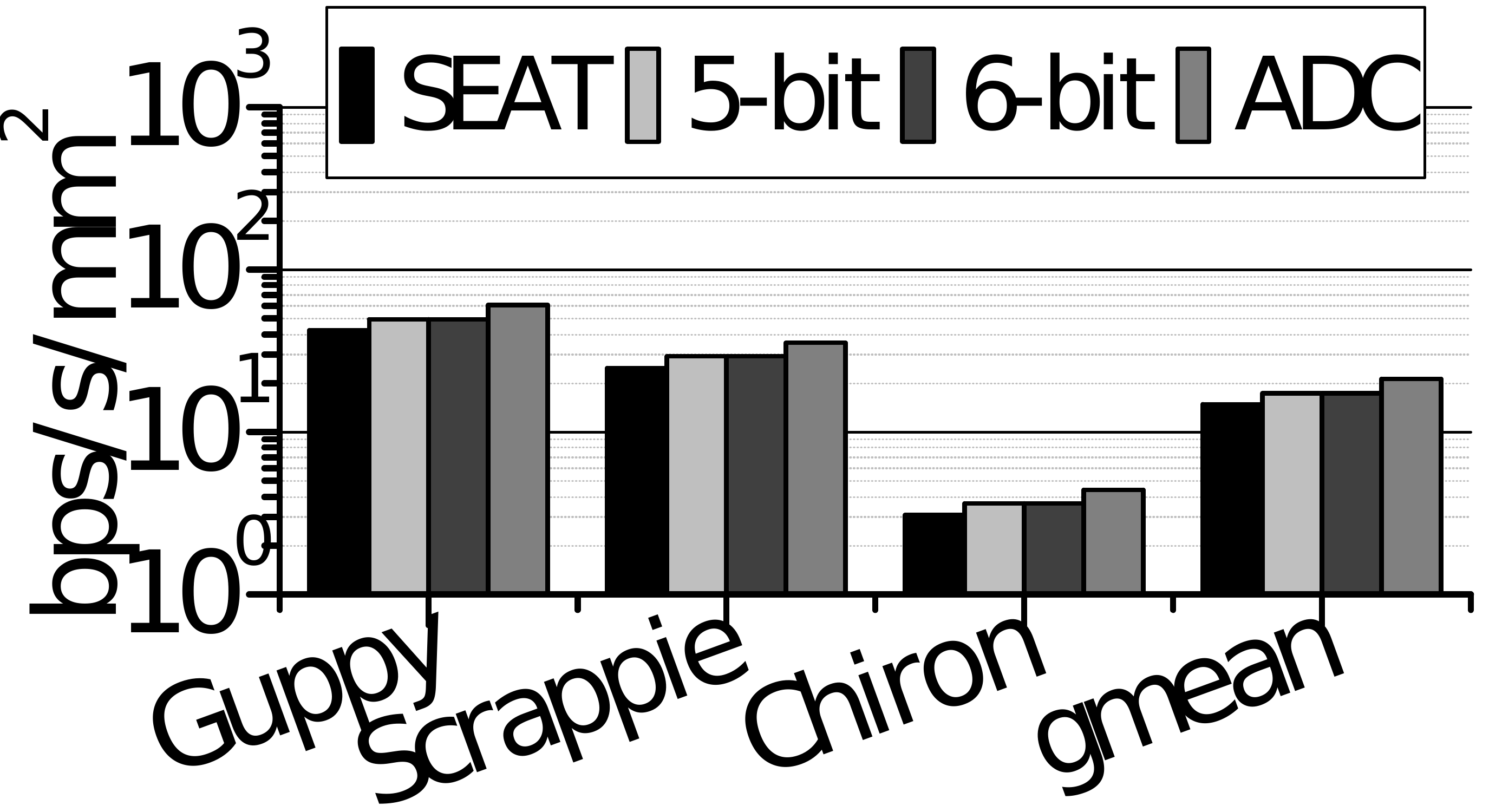}
\label{f:dna_adc_mm}
}
\vspace{-0.1in}
\caption{The comparison against various CMOS ADCs.}
\label{f:dna_adc_kmm}
\vspace{-0.1in}
\end{figure}

\subsection{ADC-free PIM Accelerator} 

\textbf{Performance per Watt and per $mm^2$}. Because of SEAT, base-callers can be quantized with 5-bit without accuracy loss. In this way, we can use our SOT-MRAM-based ADC arrays with lower resolution to reduce power consumption and area overhead of our PIM accelerator. After we replace the CMOS ADCs in \texttt{SEAT} by our SOT-MRAM-based ADC arrays (\texttt{ADC}), the PIM accelerator running 5-bit quantized base-callers can still achieve the same performance as \texttt{SEAT}, as shown in Figure~\ref{f:dna_perf_all}. However, \texttt{ADC} significantly reduces power consumption and area overhead of the PIM accelerator. As Figure~\ref{f:dna_perfwatt_all} shows, on average, \texttt{ADC} improves base-calling throughput per Watt by 127\% over \texttt{SEAT}. Moreover, \texttt{ADC} increases base-calling throughput per $mm^2$ by 42.9\%, as shown in Figure~\ref{f:dna_mm_all}.

\textbf{Comparison against ADCs with lower resolution}. Recent works rely on CMOS ADCs with lower resolutions, e.g., 5-bit~\cite{Fujiki:ASPLOS2018} and 6-bit~\cite{Yang:ISCA2019}, to reduce power consumption and area overhead of NVM-based dot-product engines. The lower resolution a CMOS ADC achieves, the smaller power consumption and area overhead it costs. We showed the comparison of performance per Watt and per $mm^2$ between NVM-based dot-product engines with our ADC arrays and with low-resolution CMOS ADCs in Figure~\ref{f:dna_adc_kmm}. As Figure~\ref{f:dna_adc_perf} shows, on average, our ADC arrays improve base-calling throughput per Watt by 27.9\% and 37.3\% over 5-bit and 6-bit CMOS ADCs respectively. Furthermore, on average, our ADC arrays increase base-calling throughput per $mm^2$ by 21.8\% and 21.3\% over 5-bit and 6-bit CMOS ADCs respectively, as shown in Figure~\ref{f:dna_adc_mm}. This is because a 5-bit CMOS ADC has similar area overhead to that of a 6-bit CMOS ADC.

\subsection{CTC Decoding and Read Vote}

\textbf{CTC decoding}. After we processed CTC decoding operations by NVM-based dot-product engines, as Figure~\ref{f:dna_perf_all} show, on average, \texttt{CTC} improves base-calling throughput by 67.8\% over \texttt{ADC}. Particularly, \texttt{CTC} boosts base-calling throughput of Chiron to $2.74\times$. Moreover, \texttt{CTC} also reduces the data transfers between the GPU and our PIM accelerator. In \texttt{CTC}, CTC decoding operations and DNN inferences share the same NVM-based dot-product engines, so \texttt{CTC} does not increase power consumption or area overhead. As a result, \texttt{CTC} improves base-calling throughput per Watt and per $mm^2$ by 64\% and 69\% over \texttt{ADC} respectively, as shown in Figure~\ref{f:dna_perfwatt_all} and~\ref{f:dna_mm_all}.

\begin{figure}[t!]
\centering
\subfigure[Performance/Watt]{
\includegraphics[width=1.55in]{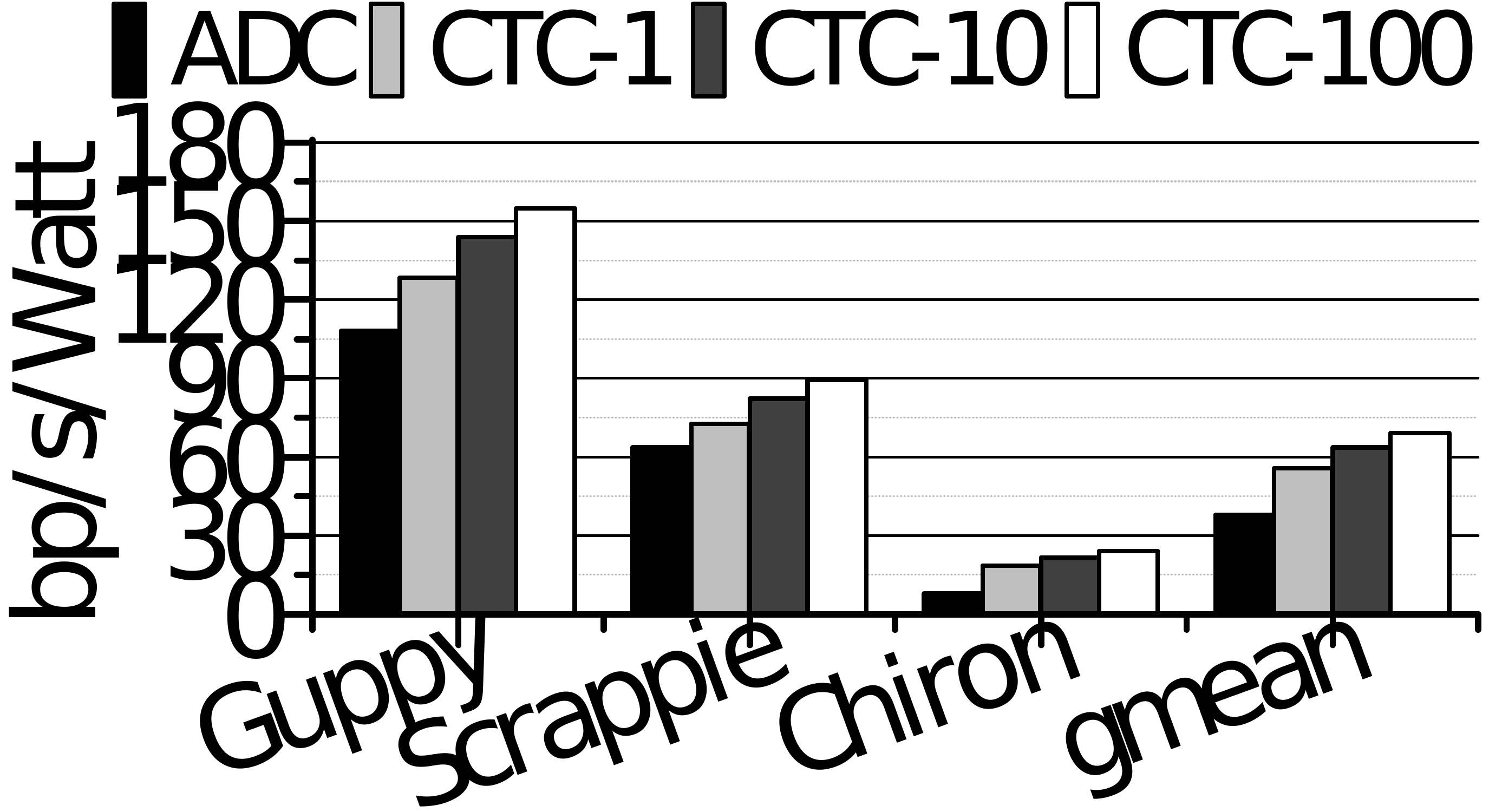}
\label{f:dna_beam_width}
}
\subfigure[Performance/$mm^2$]{
\includegraphics[width=1.55in]{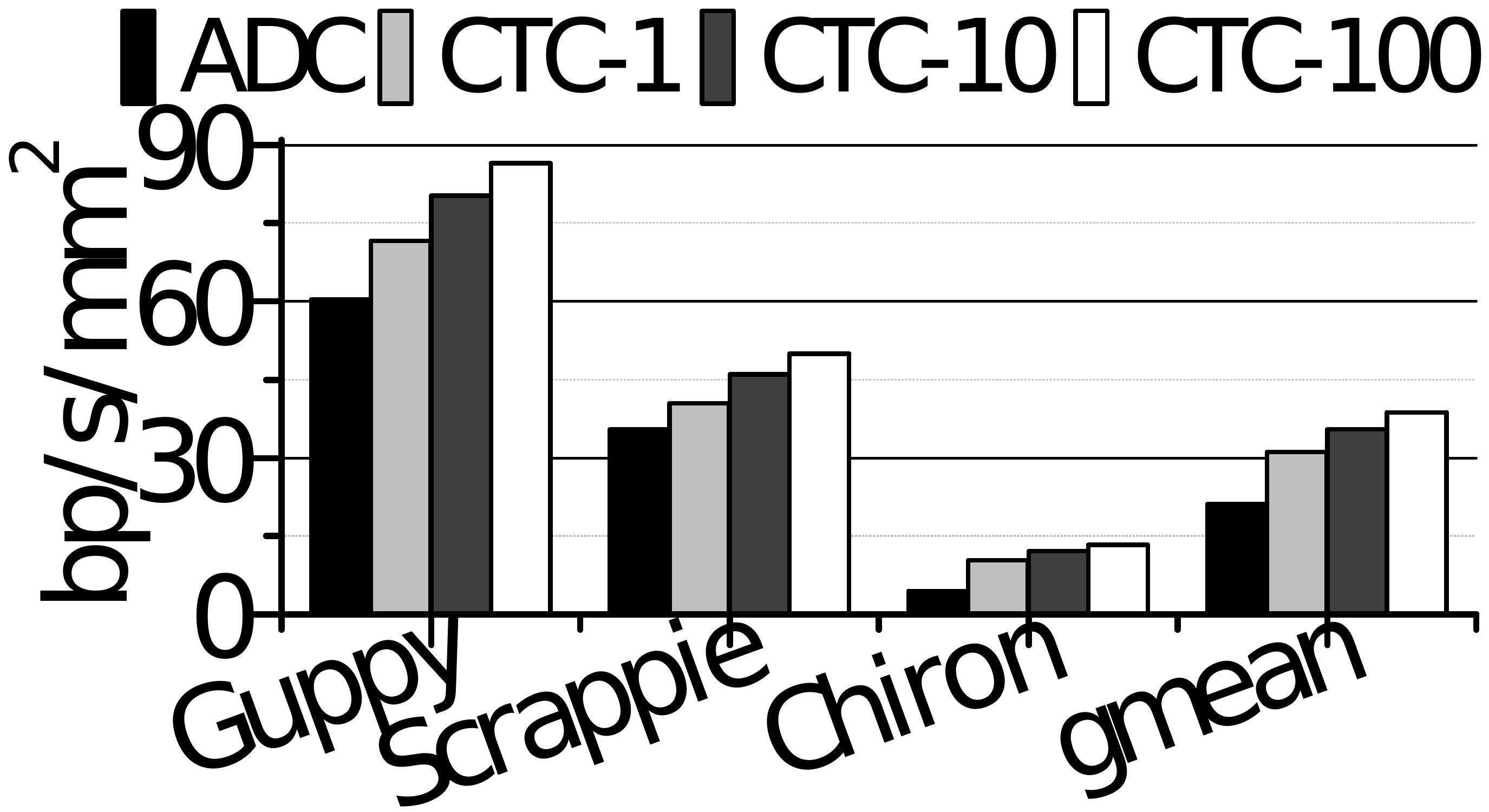}
\label{f:dna_beam_length}
}
\vspace{-0.1in}
\caption{The comparison w. varying beam search widths.}
\label{f:dna_beam_length222}
\vspace{-0.2in}
\end{figure}

\textbf{Sensitivity to beam search width}. Figure~\ref{f:dna_beam_length222} exhibits the sensitivity of base-calling throughput of \texttt{CTC} with varying beam search widths. With an enlarging width of beam search in the CTC decoder, \texttt{CTC} achieves larger improvement on base-calling throughput per Watt and per $mm^2$. This is because, with a larger width of beam search in the CTC decoder, the execution time of CTC decoding operations becomes more and more significant. A NVM-based dot-product engine requires more iterations to process a CTC decoding operation with larger beam search width.

\textbf{Read voting}. By enabling SOT-MRAM-based binary comparator arrays to process read votes, we have all proposed techniques for \texttt{Helix}. On average, \texttt{Helix} improves base-calling throughput by $2.22\times$ over \texttt{CTC}, as shown in Figure~\ref{f:dna_perf_all}. \texttt{Helix} can concurrently compare up to 256 reads by only one binary comparator array during each read voting without introducing significant power consumption or area overhead. As Figure~\ref{f:dna_perfwatt_all} and~\ref{f:dna_mm_all} show, \texttt{Helix} boosts base-calling throughput per Watt and per $mm^2$ to $3.06\times$ and $3.22\times$ over \texttt{CTC}, respectively. Overall, on average, \texttt{Helix} achieves $6\times$ base-calling throughput of \texttt{ISAAC}.

\section{Related Work}
\label{s:related}

\textbf{Nanopore sequencing}. Nanopore sequencing~\cite{Jain:Nature2018} emerges as one of the most promising genome sequencing technologies to enabling personalized medicine, global food security, and virus surveillance, because its capability of generating long reads and good real-time mobility. In a nanopore sequencing pipeline, the step of base-calling costs 44.5\% of total execution time, because of high computing overhead of state-of-the-art DNN-based base-callers. It takes more than one day for a server-level GPU to base-call a 3G-bp human genome with a $30\times$ coverage by a DNN-based base-caller. This is unacceptably slow particularly during virus outbreaks.

\textbf{Network quantization}. Although prior works propose network quantization~\cite{Li:CVPR2019,Xu:ICLR2018,Lin:ICML2016} to approximate floating-point network parameters by fixed-point representations with lower bit-widths, na\"ively applying prior network quantization on base-callers greatly increased the number of systematic errors that cannot be corrected by read votes, thereby substantially degrading the quality of final genome mappings.

\textbf{NVM dot-product engines}. Although ReRAM-~\cite{Shafiee:ISCA2016,Fujiki:ASPLOS2018,Yang:ISCA2019}, PCM-~\cite{Ambrogio:IEDM2019} , and STT-MRAM~\cite{Yan:ICS2018}-based dot-product engines are proposed in order to accelerate DNN inferences, their power efficiency and scalability are limited by power-hungry CMOS ADCs. CMOS ADCs cost 58\% of power consumption and 30\% of chip area in a well-known ReRAM-based PIM~\cite{Shafiee:ISCA2016}. Another recent ReRAM-based PIM~\cite{Fujiki:ASPLOS2018} consumes $416W$ and has power density of $842mW/mm^2$, much larger than the thermal tolerance of a ReRAM chip with active heat sinks~\cite{Zhu:MEMSYS2016}.

\textbf{Hardware acceleration for genome sequencing}. Hardware specialized acceleration is an effective way to overcome the big genomic data problem. However, most prior works focus on only accelerating genome alignment and assembly~\cite{Turakhia:ASPLOS2018}, particular short read alignment~\cite{Wu:HPCA2019,Turakhia:HPCA2019,Madhavan:ISCA2014,Fuijiki:ISCA2018,Huangfu:MICRO2019,Zokaee:PACT2019}. However, long read alignment and assembly are not the most-time consuming steps in a nanopore sequencing pipeline.

\section{Conclusion}
\label{s:con}

In this paper, we proposed an algorithm/architecture co-designed PIM accelerator, Helix, to process nanopore base-calling. We presented systematic error aware training to decrease the bit-width of a quantized base-caller without increasing the number of systematic errors that cannot be corrected through read voting operations. We also create a SOT-MRAM ADC array to accelerate analog-to-digital conversion operations. Finally, we revised a traditional NVM-based dot-product engine to accelerate CTC decoding operations, and then introduced a SOT-MRAM binary comparator array to process read voting operations at the end of base-calling. Compared to state-of-the-art PIM accelerators, Helix improves base-calling throughput by $6\times$, throughput per Watt by $11.9\times$, and per $mm^2$ by $7.5\times$ without degrading base-calling accuracy.

\begin{acks}
We thank the anonymous reviewers for their insightful comments and constructive suggestions. This work was supported in part by NSF CCF-1909509 and CCF-1908992.
\end{acks}

\balance
\bibliographystyle{ACM-Reference-Format}
\bibliography{genomics}

\end{document}